\documentclass[]{jfm}

\usepackage[english]{babel}
\usepackage{epsfig}
\usepackage{amsmath}
\usepackage{amssymb}
\usepackage{comment}
\usepackage{graphicx}
\usepackage{nicefrac}
\usepackage{natbib}

\usepackage{srcltx} 

\newcommand{\bs}{\boldsymbol}
\newcommand{\mat}{\mathrm}

\begin{document}

\title[On the velocity distribution in homogeneous isotropic turbulence]{On the velocity distribution in homogeneous isotropic turbulence: correlations and deviations from Gaussianity}

\author[M. Wilczek, A. Daitche, R. Friedrich]{Michael Wilczek, Anton Daitche, Rudolf Friedrich}
\affiliation{Institute for Theoretical Physics, Westf\"alische
  Wilhelms-Universit\"at, Wilhelm-Klemm-Stra\ss e\,9, 48149 M\"unster, Germany}

\date{\today}

\maketitle

\begin{abstract}
We investigate the single-point probability density function of the velocity in three-dimensional stationary and decaying homogeneous isotropic turbulence. To this end we apply the statistical framework of the Lundgren--Monin--Novikov hierarchy combined with conditional averaging, identifying the quantities that determine the shape of the probability density function. In this framework the conditional averages of the rate of energy dissipation, the velocity diffusion and the pressure gradient with respect to velocity play a key role. Direct numerical simulations of the Navier--Stokes equation are used to complement the theoretical results and assess deviations from Gaussianity.
\end{abstract}

\section{Introduction}
The spatio-temporal complexity of turbulent flows demands a statistical description, which can be formulated in terms of probability density functions (PDFs) of the fluctuating turbulent quantities such as velocity or vorticity. The Lundgren--Monin--Novikov (LMN) hierarchy \cite[]{lundgren67pof,monin67pmm,novikov68sdp,ulinich69spj} provides a statistical framework, which, starting from the basic equations of motion (i.e. the Navier--Stokes equation or the vorticity equation), yields the temporal evolution of the multi-point PDFs of the velocity or vorticity. As typical for a statistical theory of turbulence, this approach has to face the famous closure problem of turbulence. When deriving the evolution equation of the single-point PDF, this equation couples to the two-point PDF. Now determining the evolution of the two-point PDF involves the three-point PDF and so forth, ending up in an ever increasing number of evolution equations for the multi-point PDFs. This chain of equations may be truncated at a given level by introducing reasonable closure approximations, which express the $(n+1)$-point PDF in terms of the $n$-point PDF. However, these approximations have to be performed with great care as a too crude approximation results in serious defects. An alternative approach is provided by conditional averaging, which, for example, is taken as a starting point for modelling in the realm of the well-known PDF methods by \cite{pope00book}. \cite{novikov93jfr} used this method in the case of the vorticity equation to elucidate the statistical balance of vortex stretching and vorticity diffusion. When combined with the LMN hierarchy, a truncation at a given level is possible by introducing the conditional averages of the terms appearing in the equations of motion as unknown functions. These functions then may be modelled or estimated from experiments or direct numerical simulations as recently exemplified in \cite[]{wilczek09pre} in the case of the single-point vorticity PDF. The advantage of this approach lies in the possibility to interpret the unknown functions in terms of their physical meaning. For instance, it turns out that the single-point velocity PDF depends on the conditionally averaged rate of energy dissipation.\par 
Ever since the experimental measurements by Townsend presented in \cite[]{batchelor53book} it is an ongoing discussion, whether the single-point velocity PDF displays a Gaussian shape. Arguments in favour of this hypothesis often involve the central limit theorem, taking the velocity field as a spatial ensemble of independently fluctuating random variables with finite variances. This argument turns out to be rather hand-waving as the assumption of statistical independence is obviously violated. Moreover, the same kind of argument should be applicable to the vorticity PDF. This PDF, however, is known to display extremely non-Gaussian tails, which are often associated with the existence of vortex structures in the flow. When applied with care, however, the central limit theorem together with proper assumptions on the energy spectrum has been shown to produce sub-Gaussian tails by \cite{jimenez97jfm}.\par
While the assumption of Gaussianity at least seems to be a good approximation, experimental and numerical measurements indicate the possibility of sub-Gaussian tails \cite[]{vincent91jfm,noullez97jfm,gotoh02pof}. Theoretical support in this direction was, in the case of forced turbulence, given by \cite{falkovich97prl}. The approach, however, depends on the type of forcing applied to the flow. This point recently was raised in \cite[]{hosokawa08pre}, where it was hypothesized that decaying turbulence might be profoundly different from forced turbulence. In this work, a closure for the LMN hierarchy was suggested leading to a Gaussian velocity PDF for decaying turbulence. Gaussian PDFs are also found as a result of the cross-independence hypothesis  \cite[]{tatsumi04fdr} for both the single-point PDF and joint PDFs in the inertial range. In both works the pressure contributions are neglected on the level of the single-point PDF.\par 
The present work aims at clarifying these points. We make use of the statistical framework of the LMN hierarchy combined with conditional averaging to identify the terms which determine the shape and evolution of the single-point velocity PDF. This approach gives concise insights how the different terms interplay to determine the functional form of the PDF. The theory applies to forced as well as decaying turbulence. The results presented in \cite[]{hosokawa08pre} can be obtained via a simple approximation, which neglects the statistical dependence of the pressure gradient and the energy dissipation on the velocity. Testing the theory against results from direct numerical simulation, however, indicates slightly non-Gaussian PDFs for both forced and decaying turbulence, which is related to the statistical correlations between the velocity and the different dynamical effects such as pressure gradient and energy dissipation. \par
The remainder of this article is structured as follows. After reviewing the LMN hierarchy and conditional averaging, we discuss the implications of statistical symmetries, which help to drastically simplify the description. We then derive explicit formulae for the homogeneous and stationary velocity PDF. The conditional averages appearing in these formulae have to obey certain constraints which will be derived and used to motivate a simple closure approximation. We then turn to the numerical results of stationary and decaying turbulence examining the different quantities arising in the theoretical description. To highlight the genuine properties of the velocity statistics, some comparisons with the single-point statistics of the vorticity are drawn before we conclude.

\section{Theoretical Framework}

\subsection{The Lundgren--Monin--Novikov Hierarchy}
As an introduction into the statistical framework used in this work, we start with revisiting the derivation of the LMN hierarchy. For further details, we refer the reader to \cite[]{lundgren67pof,monin67pmm,novikov68sdp,pope00book}.\par
The Navier--Stokes equation for an incompressible fluid reads as
\begin{equation}\label{eq:navier-stokes}
  \frac{\partial}{\partial t}\bs u(\bs x,t)+\bs u(\bs x,t)\cdot\bs{\nabla} \bs u(\bs x,t)=-\bs{\nabla} p(\bs x,t)+\nu \Delta \bs u(\bs x,t)+\bs F(\bs x,t) .
\end{equation}
Here $\bs u$ denotes the velocity field, $p$ is the pressure, $\nu$ denotes the kinematic viscosity and $\bs F$ denotes a large-scale forcing applied to the fluid to produce a statistically stationary flow. In the case of $\bs F(\bs x,t)=\bs 0$, turbulence decays due to dissipation and statistical stationarity cannot be maintained. \par
Let $\bs u(\bs x,t)$ denote a realization of the velocity field and $\bs v$ denote the corresponding sample space variable. The fine-grained PDF of the velocity is then defined by $\hat f(\bs v;\bs x,t)=\delta(\bs u(\bs x,t)-\bs v)$, from which the PDF can be obtained by ensemble-averaging,
\begin{equation}
  f(\bs v;\bs x,t)=\langle \hat f(\bs v;\bs x,t) \rangle=\langle \delta(\bs u(\bs x,t)-\bs v) \rangle .
\end{equation}
Taking the derivative of the fine-grained PDF with respect to time yields
\begin{equation}
  \frac{\partial }{\partial t} \hat f(\bs v;\bs x,t)=-\bs{\nabla}_{\bs v} \cdot \bigg[ \frac{\partial \bs u}{\partial t}(\bs x,t) \, \hat f(\bs v;\bs x,t) \bigg] .
\end{equation}
Together with an analogous calculation for the advective term we obtain
\begin{align}\label{eq:finegrained}
 \frac{\partial }{\partial t} \hat f+\bs{\nabla}\cdot \left[ \bs u \hat f \right] &= -\bs{\nabla}_{\bs v} \cdot \bigg[ \left( \frac{\partial \bs u}{\partial t}+\bs u\cdot\bs{\nabla} \bs u \right) \hat f \bigg] \nonumber\\
  &= -\bs{\nabla}_{\bs v} \cdot \bigg[ \left( -\bs{\nabla} p+\nu \Delta \bs u + \bs F \right) \hat f \bigg] .
\end{align}
Here, the Navier--Stokes equation \eqref{eq:navier-stokes} has been used to replace the terms of the left-hand side with the terms of the right-hand side. Incompressibility ($\bs{\nabla}\cdot\bs u=0$) was used to rearrange the advective term. In order to turn this into an equation for the PDF, we have to perform an ensemble average of \eqref{eq:finegrained}. The averaging of the left-hand side is straightforward yielding $\partial_t f+\bs v \cdot \bs{\nabla} f$; however, the terms on the right-hand side cannot be expressed in terms of $f$ and $\bs v$ only. Now there are several options how to proceed. One possibility is to express these unclosed terms on the right-hand side with the help of the two-point PDF. We exemplify this for the pressure term.\par
The two-point PDF of the velocity can be written as $f_2(\bs v_1,\bs v_2; \bs x_1,\bs x_2,t)=\langle \delta(\bs u(\bs x_1,t)-\bs v_1) \, \delta(\bs u(\bs x_2,t)-\bs v_2) \rangle$. Now the pressure term can be expressed in terms of the velocity according to
\begin{align}
  \big\langle \big( -\bs{\nabla}_{\bs x_1} p \big) \hat f  \big\rangle
  &= \bigg\langle -\frac{1}{4\pi}\int\! \mathrm{d}\bs x_2 \left(\bs{\nabla}_{\bs x_1}\frac{\bs{\nabla}_{\bs x_2} \cdot [ \bs u(\bs x_2,t)\cdot\bs{\nabla}_{\bs x_2}\bs u(\bs x_2,t) ]}{|\bs x_2-\bs x_1|} \right) \, \delta(\bs u(\bs x_1,t)-\bs v_1) \bigg\rangle \displaybreak[0] \nonumber \\
  &= \bigg\langle -\frac{1}{4\pi} \int\! \mathrm{d}\bs x_2 \left( \bs{\nabla}_{\bs x_1} \frac{\mathrm{Tr}\,[(\bs{\nabla}_{\bs x_2}\bs{\nabla}_{\bs x_2}^{\mathrm T}) (\bs u(\bs x_2,t) \bs u^{\mathrm T}(\bs x_2,t))]}{|\bs x_2-\bs x_1|} \right) \,  \times \nonumber  \\
  & \qquad \qquad \qquad \delta(\bs u(\bs x_1,t)-\bs v_1) \left( \int\! \mathrm{d}\bs v_2 \, \delta(\bs u(\bs x_2,t)-\bs v_2) \right) \bigg \rangle \nonumber  \displaybreak[0]\\
  &= -\frac{1}{4\pi} \int\! \mathrm{d}\bs x_2 \, \mathrm{d}\bs v_2 \left( \bs{\nabla}_{\bs x_1} \frac{1}{|\bs x_2-\bs x_1|} \right) \, \left( \bs v_2 \cdot \bs{\nabla}_{\bs x_2} \right)^2 \, f_2(\bs v_1,\bs v_2; \bs x_1,\bs x_2,t) ,
\end{align}
where we have assumed an infinite domain without further boundary conditions. Here, the identity \mbox{$\int \mathrm{d}\bs v_2 \, \delta(\bs u(\bs x_2,t)-\bs v_2)=1 $}, incompressibility and the sifting property of the delta function, $\bs u(\bs x_2,t) \, \delta(\bs u(\bs x_2,t)-\bs v_2)=\bs v_2  \, \delta(\bs u(\bs x_2,t)-\bs v_2)$, have been used. This manipulation explicitly shows how the single-point statistics couples to the two-point statistics. The remaining unclosed terms can be treated in a similar manner; for more details, we refer the reader to \cite[]{lundgren67pof}. Now determining the evolution equation for $f_2$ basically involves the same steps, leading to a coupling to $f_3$. This eventually leads to the aforementioned never-ending hierarchy of evolution equations for the multi-point velocity distributions.\par
Instead of establishing this chain of evolution equations, it is possible to truncate the hierarchy on a given level by introducing conditional averages according to e.g.
\begin{equation}
 \big\langle (-\bs{\nabla} p) \hat f  \big\rangle=\big\langle (\, -\bs{\nabla} p(\bs x,t) \,) \, \delta(\bs u(\bs x,t) - \bs v)  \big\rangle=\big\langle -\bs{\nabla} p \big| \bs v  \big\rangle f,
\end{equation}
i.e. now the coupling to a higher-order PDF is replaced by introducing unknown functions in the form of conditionally averaged quantities, which may be modelled or obtained by experimental or numerical means. The conditional averages may be regarded as a measure for the correlation (in the sense of statistical dependence) of the different quantities on the single-point level. We will focus on this approach in the following. The temporal evolution of the velocity PDF then takes the general form
\begin{equation}\label{eq:pdfvel}
  \frac{\partial}{\partial t}f+\bs v\cdot\bs{\nabla} f=-\bs{\nabla}_{\bs v} \cdot \left[ \big\langle -\bs{\nabla} p+\nu\Delta \bs u + \bs F \big| \bs v \big\rangle f \right].
\end{equation}
To rewrite the advective term, here again incompressibility and the sifting property have been used. In this kinetic description for the evolution of the single-point velocity PDF, several unclosed terms appear: the conditionally averaged pressure gradient, the Laplacian of the velocity as well as the external forcing. We take this equation as a starting point for the following considerations and will now utilize statistical symmetries to simplify the description.

\subsection{Statistical Symmetries}
\subsubsection{Homogeneity}
For homogeneous flows, several simplifications arise. First, the advective term on the left-hand side of \eqref{eq:pdfvel} vanishes as homogeneity means that $f$ cannot depend on $\bs x$. This simplifies the kinetic equation resulting in
\begin{equation}\label{eq:pdfvelhomo}
  \frac{\partial}{\partial t}f=-\bs{\nabla}_{\bs v} \cdot \left[ \big\langle -\bs{\nabla} p + \nu\Delta \bs u + \bs F \big| \bs v \big\rangle f \right].
\end{equation}
Note that in the case of decaying turbulence, the evolution is determined by the conditionally averaged pressure gradient and the dissipative term only. Homogeneity allows to recast the latter. Calculating the Laplacian of $\hat f$ and averaging yields\footnote{Throughout the paper the Einstein summation convention is assumed.}
\begin{equation}\label{eq:hom}
  \frac{\partial^2}{\partial x_k^2} f = 0 =-\frac{\partial }{\partial v_i}  \bigg \langle  \frac{\partial^2 u_i}{\partial x_k^2}  \bigg | \bs v  \bigg \rangle f +\frac{\partial }{\partial v_i}\frac{\partial }{\partial v_j}  \bigg \langle \frac{\partial u_i}{\partial x_k} \frac{\partial u_j}{\partial x_k} \bigg | \bs v  \bigg \rangle f.
\end{equation}
This is a multi-dimensional version of the homogeneity relation introduced by \cite{ching96pre}. With this expression, the kinetic equation takes the form
\begin{equation}\label{eq:pdfvelhomo2}
  \frac{\partial}{\partial t}f=-\frac{\partial}{\partial v_i} \bigg \langle -\frac{\partial}{\partial x_i} p +  F_i  \bigg | \bs v  \bigg \rangle f - \frac{\partial }{\partial v_i}\frac{\partial }{\partial v_j} \bigg \langle \nu \frac{\partial u_i}{\partial x_k} \frac{\partial u_j}{\partial x_k}  \bigg | \bs v  \bigg \rangle f ,
\end{equation}
which has also been derived for turbulence with mean flow by \cite{pope00book}.
This introduces the tensor $D_{ij}= \big \langle  \nu (\partial u_i/\partial x_k) (\partial u_j/\partial x_k) \big | \bs v  \big \rangle $, sometimes termed conditionally averaged dissipation tensor, which will be of central interest in the following.\par
Before we proceed to the simplifications that arise due to isotropy, we pause for a side remark as we would like to highlight the connection to moment equations which can be obtained directly from the PDF equation \eqref{eq:pdfvelhomo2}. The law of energy decay may serve as an example. In the decaying case, the only two terms in \eqref{eq:pdfvelhomo2} governing the evolution of the velocity PDF are the conditionally averaged pressure gradient and the conditional dissipation tensor. While the former takes the form of a drift induced by the non-local pressure contributions, the latter term may be interpreted as a diffusive term, however with a negative sign. This sign can be explained in a physically sound way as for a diffusion process an initially localized concentration spreads over time, eventually being dispersed over a large domain. Regarding the decay of the velocity field of a turbulent fluid, the opposite takes place. An initially broad distribution of velocity contracts as the velocity field dies away. When the fluid has come to rest, the PDF is localized sharply in probability space, $ \lim \limits_{ t \rightarrow \infty} f(\bs v;t)=\delta(\bs v)$, expressing that we have probability one finding a vanishing velocity.\par
To proceed, we multiply \eqref{eq:pdfvelhomo2} by $\bs v^2/2$ and integrate over $\bs v$,
\begin{equation}
  \frac{\partial}{\partial t} E_{kin}=\int \mathrm{d}\bs v \frac{\bs v^2}{2}  \frac{\partial}{\partial t}f=-\int \mathrm{d}\bs v \frac{v_k v_k}{2} \left[  \frac{\partial }{\partial v_i} \bigg \langle -\frac{\partial}{\partial x_i} p \bigg | \bs v \bigg \rangle f + \frac{\partial }{\partial v_i}\frac{\partial }{\partial v_j} \bigg\langle \nu \frac{\partial u_i}{\partial x_k} \frac{\partial u_j}{\partial x_k} \bigg | \bs v \bigg\rangle f \right ].
\end{equation}
The right-hand side may be integrated by parts. Assuming a sufficiently rapid decay of the PDF, one finds
\begin{align}\label{eq:endecay}
  \frac{\partial}{\partial t} E_{kin} &=  \int \mathrm{d}\bs v \, \bigg [ -\big\langle \bs u \cdot \bs{\nabla} p \big| \bs v \big\rangle f -\bigg\langle \nu \frac{\partial u_i}{\partial x_k} \frac{\partial u_i}{\partial x_k} \bigg | \bs v \bigg\rangle f\bigg] \nonumber\\
                       &= -\big\langle \bs u \cdot \bs{\nabla} p\big\rangle -\frac{1}{2} \big\langle \varepsilon + \nu \omega^2 \big\rangle \nonumber\\
                       &= - \big\langle \varepsilon \big\rangle .
\end{align}
The validity of the second equality will become clear in the next section (see \eqref{eq:trace}). The last equality comes from the fact that the pressure-related average vanishes due to homogeneity and incompressibility and that the rate of energy dissipation and squared vorticity (multiplied by $\nu$) have the same spatial and hence also ensemble average. Equation \eqref{eq:endecay} is what one expects and also what is long known as the evolution equation of the kinetic energy. It is not surprising that the correct equation for the energy follows from the equations for the velocity PDF as those were derived directly from the Navier--Stokes equation without using any approximations.

\subsubsection{Isotropy}
The conditional averages appearing in \eqref{eq:pdfvelhomo} and \eqref{eq:pdfvelhomo2} are vector- and symmetric-tensor-valued functions of the velocity vector $\bs v$. Statistical isotropy, which is here assumed as invariance under rotation\footnote{Invariance under reflections is not necessary for the results presented in this paper.}, imposes further constraints on these functions e.g.
\begin{equation}
  \mat{D}(\mat{R}\bs v)=\mat{R}\mat{D}(\bs v)\mat{R}^{\mathrm T}\qquad \mat{R}\in\mathrm{SO}(3),
  \label{}
\end{equation}
where $\mat{R}$ is a rotation matrix.
In three dimensions it can be shown that, because of these constraints, isotropic vector- and symmetric-tensor-valued functions like $\langle\bs{\nabla} p\big |\bs v\rangle$ and $\mat{D}(\bs v)$ take the form \cite[]{robertson40pps,batchelor53book}
\begin{align}
   a_i(\bs v) &= a(v) \, \frac{v_i}{v} \label{eq:isotropy_a},\\
   B_{ij}(\bs v) &= \mu(v) \, \delta_{ij} + \left[ \lambda(v)-\mu(v)\right]  \frac{v_i v_j}{v^2} \label{eq:isotropy_B},
\end{align}
where $a(v)=\hat{\bs v}\cdot\bs a(\bs v)$ is the projection of $\bs a(\bs v)$ onto the unit vector in the direction of the velocity and $\lambda(v)$, $\mu(v)$ are the eigenvalues of the matrix $\mat{B}$. Here $\lambda$ is the eigenvalue of the eigenvector $\bs v$, whereas $\mu$ is the eigenvalue of the (two-dimensional) eigenspace perpendicular to $\bs v$. Note that because $a$, $\lambda$ and $\mu$ are isotropic scalar-valued functions, they depend only on the absolute value of the velocity $v$. The same goes for the probability density $f(\bs v)$, i.e. it is only a function of the absolute value $v$. This, however, may not be confused with the PDF of the absolute value of the velocity, in the following denoted by $\tilde f(v)$. However, there is a simple relation between those two,
\begin{equation}\label{eq:isopdf}
 \tilde f(v) = 4 \pi v^2 f(\bs v),
\end{equation}
indicating that it suffices to determine the PDF of the absolute value in order to specify the PDF of the full vector. 

From the consideration about isotropic vector-valued functions, we find
\begin{align}
 \left\langle -\bs{\nabla} p \big | \bs v \right\rangle  = \Pi(v) \hat{\bs v}, 
 &\qquad \Pi(v)=\left\langle - \hat{\bs u} \cdot \bs{\nabla} p \big | v \right\rangle \label{eq:iso-pressure},\\ 
 \left\langle \nu \Delta \bs u \big | \bs v \right\rangle = \Lambda(v) \hat{\bs v}, 
&\qquad \Lambda(v)=\left\langle \nu  \hat{\bs u} \cdot \Delta \bs u \big | v \right\rangle \label{eq:iso-laplace},\\
 \left\langle \bs F \big |  \bs v \right\rangle = \Phi(v) \hat{\bs v}, 
&\qquad \Phi(v)=\left\langle \hat{\bs u} \cdot \bs F \big | v \right\rangle \label{eq:iso-force},
\end{align}
where $\hat{\bs u}$ and $\hat{\bs v}$ denote the unit vectors in the direction of $\bs u$ and $\bs v$. Note that it is sufficient to take the conditional average with respect to the magnitude of the velocity to obtain the functions $\Pi$, $\Lambda$ and $\Phi$. This considerably simplifies the numerical estimation of the conditional averages. Let us now turn to the conditional dissipation tensor, which takes the form
\begin{equation}
  \mat{D}(\bs v)=\big\langle \nu \mat{A}\mat{A}^{\mathrm T} \big| \bs v \big\rangle,
\end{equation}
where $A_{ij}=\partial u_i/\partial x_j$ is the velocity gradient tensor. $\mat{A}$ may be decomposed into symmetric and antisymmetric parts according to $\mat{A}=\mat{S}+\mat{W}$, where $\mat{S}=\frac{1}{2}(\mat{A}+\mat{A}^{\mathrm T})$ and $\mat{W}=\frac{1}{2}(\mat{A}-\mat{A}^{\mathrm T})$. These two tensors characterize the local rate of stretching and the rate of rotation of the fluid. In the case of statistical isotropy, $\mat{D}$ is determined by its eigenvalues (see \eqref{eq:isotropy_B}), which can be obtained using the relations
\begin{align}
  \mathrm{Tr}(\mat{D}) & =\lambda(v) + 2\mu(v),\\
\hat{\bs v}\mat{D}\hat{\bs v} &=\lambda(v) .
\end{align}
The trace of $\mat{D}$ is determined by the conditional averages of the local rate of energy dissipation $\varepsilon=2\nu \mathrm{Tr}(\mat{S}^2)$ and the squared vorticity $\omega^2$:
\begin{equation}\label{eq:trace}
  \mathrm{Tr}(\mat{D})=\nu \left\langle   \mathrm{Tr}(\mat{S}^2)-  \mathrm{Tr}(\mat{W}^2) \big | v \right\rangle =\frac{1}{2}\left\langle  \varepsilon+\nu \omega^2 \big | v\right\rangle .
\end{equation}
The second scalar quantity needed to determine $\mat{D}$ is
\begin{equation}\label{eq:con2}
 \hat{\bs v}\mat{D}\hat{\bs v}=\left\langle \nu \, \hat{\bs u}\mat{AA}^{\mathrm T}\hat{\bs u}\big | v\right\rangle =\left\langle \nu(\mat{A}^{\mathrm T}\hat{\bs u})^2\big | v\right\rangle .
\end{equation}
This rather formally looking quantity has a simple physical interpretation. As $\mat{A}$ may be decomposed in symmetric and antisymmetric parts, we write $\mat{A}^{\mathrm T} \hat{\bs  u }=(\mat{S}-\mat{W})\hat{\bs u }$. The last term may also be written as $\mat{W}\hat{\bs u}=\frac{1}{2}\bs \omega \times \hat{\bs u}$ due to the relation $W_{ij}=-\frac{1}{2}\epsilon_{ijk}\omega_k$. Hence, the conditional average appearing in \eqref{eq:con2} involves the absolute value of the difference between the rate of stretching in the direction of the velocity vector and rate of rotation of the unit vector $\hat{\bs u}$.
Summing up, the conditional dissipation tensor $\mat{D}$ in isotropic turbulence has the form
\begin{align}
  D_{ij}(\bs v) &= \mu(v) \, \delta_{ij} + \left[ \lambda(v)-\mu(v)\right]  \frac{v_i v_j}{v^2},\label{eq:D-lambda-mu}\\
 \mu(v) &= \frac{1}{4}\left\langle  \varepsilon+\nu \omega^2 \big | v\right\rangle-\frac{1}{2}\left\langle \nu(\mat{A}^{\mathrm T}\hat{\bs u})^2\big | v\right\rangle, \label{eq:mu_relation} \\ 
 \lambda(v) &= \left\langle \nu(\mat{A}^{\mathrm T}\hat{\bs u})^2\big | v\right\rangle \label{eq:lambda_relation} .
\end{align}
These relations can now be used to simplify the structure of the PDF equations \eqref{eq:pdfvelhomo}--\eqref{eq:pdfvelhomo2}. It turns out that
by exploiting statistical isotropy, we have to deal with an effectively one-dimensional problem. Take for example
\begin{align}
  \frac{\partial}{\partial v_i} \big\langle \nu \Delta u_i \big| \bs v \big\rangle f(\bs v) &= \frac{\partial}{\partial v_i} \Lambda(v) \frac{v_i}{4\pi v^3} \tilde f(v) \nonumber\\
  &= \frac{1}{4\pi v^2} \frac{\partial}{\partial v} \Lambda(v) \tilde f(v),
\end{align}
which makes clear that this term depends on $v$ only. The same applies to terms involving $\mat{D}$, a short calculation yields
\begin{align}
  \frac{\partial}{\partial v_i}\frac{\partial}{\partial v_j} D_{ij}(\bs v)f(\bs v)=\frac{1}{4\pi v^2} \bigg[ \frac{\partial^2}{ \partial v^2} \lambda(v) \tilde f(v)-\frac{\partial}{ \partial v} \frac{2}{v} \mu(v) \tilde f(v) \bigg].
\end{align}
This leads to the isotropic form of \eqref{eq:pdfvelhomo}--\eqref{eq:pdfvelhomo2}:
\begin{align}
\frac{\partial}{\partial t} \tilde f&=-\frac{\partial}{\partial v} \left( \Pi+\Lambda+\Phi \right) \tilde f\label{eq:pdfveliso},\\
0&=-\frac{\partial}{\partial v} \left( \Lambda+\frac{2\mu}{v} \right) \tilde f + \frac{\partial^2}{\partial v^2}  \lambda  \tilde f\label{eq:pdfhomoiso},\\
 \frac{\partial}{\partial t} \tilde f&=-\frac{\partial}{\partial v} \left( \Pi+\Phi-\frac{2\mu}{v} \right) \tilde f - \frac{\partial^2}{\partial v^2} \lambda \tilde f\label{eq:pdfvelhomoiso}.
\end{align}
We note in passing that these equations can also be derived without the assumption of statistical isotropy and therefore correctly describe $\tilde{f}(v)$ also in statistically anisotropic turbulence. However, in that case they are not equivalent to \eqref{eq:pdfvelhomo}--\eqref{eq:pdfvelhomo2} and $f(\bs v)$ is not fully determined by $\tilde{f}(v)$. Furthermore, the relations \eqref{eq:iso-pressure}--\eqref{eq:iso-force} and \eqref{eq:D-lambda-mu} are not valid in that case and do not allow a simple interpretation of the quantities $\Pi$, $\Lambda$, $\Phi$, $\lambda$ and $\mu$. 

\subsection{Homogeneous and Stationary PDFs}
Because now having reduced the problem to a one-dimensional, \eqref{eq:pdfhomoiso} is easily integrated yielding the velocity PDF in homogeneous turbulence,
\begin{equation}\label{eq:homosol}
  \tilde f(v;t)=\frac{{\cal N}}{\lambda(v,t)} \exp \int_{v_0}^v \mathrm{d}v' \, \frac{ \Lambda(v',t)+\frac{2}{v'}\mu(v',t)}{\lambda(v',t)},
\end{equation}
showing that it can be expressed as a function of the conditional averages involving the velocity diffusion, dissipation, enstrophy and the stretching and turning of the velocity vector. Here ${\cal N}$ denotes a normalization constant, which depends on $v_0$. Note that by only imposing isotropy and homogeneity, the conditional averages will be a function of time. In the following, we will refer to this solution as the `homogeneous solution'.\par
In the case of forced turbulence, it is possible to maintain a statistically stationary flow. This immediately implies $\partial_tf=0$ and from \eqref{eq:pdfveliso} follows 
\begin{equation}\label{eq:probcurrent}
\Pi(v)+\Lambda(v)+\Phi(v)=0 ,
\end{equation}
i.e. the pressure gradient term, the viscous term and the term stemming from the external forcing identically cancel. This also means that in the case of stationary, homogeneous and isotropic turbulence the probability current on the right-hand side of \eqref{eq:pdfvelhomo} identically vanishes. Note that this is not obvious and is due to the fact that the problem becomes effectively one-dimensional by virtue of isotropy. In the case of the turbulent vorticity, this balance was discussed in \cite[]{novikov93jfr,novikov94mpl} and more recently in \cite[]{wilczek09pre}.
In the stationary case, \eqref{eq:pdfhomoiso} and \eqref{eq:pdfvelhomoiso} are equivalent due to the conditional balance \eqref{eq:probcurrent}. The stationary solution of \eqref{eq:pdfvelhomoiso} reads as
\begin{equation}\label{eq:statsol}
  \tilde f(v)=\frac{{\cal N}}{\lambda(v)} \exp \int_{v_0}^v \mathrm{d}v' \, \frac{ -\Pi(v')-\Phi(v')+\frac{2}{v'}\mu(v')}{\lambda(v')},
\end{equation}
and can be either obtained by solving \eqref{eq:pdfvelhomoiso} or from \eqref{eq:homosol} by utilizing the conditional balance and omitting the time dependence. We emphasize here that introducing the second derivatives with the help of the homogeneity relation is essential to obtain a unique stationary solution which cannot be obtained solely from \eqref{eq:pdfveliso}. \par
We now have formally identified the quantities which determine the shape of the single-point velocity PDF, however, the explicit functional form of these quantities is unknown. Further input from the numerical or experimental side is needed to specify the PDF, which will be presented below, after discussing some more theoretical points. Note that the equations and relations derived so far are exact as no approximations have been made.

\subsection{Constraints on the Functional Form of the Conditional Averages}\label{sec:Constraints}
Now that the theoretical framework is set up, we are faced with the closure problem of turbulence in terms of the unknown conditional averages $\lambda$, $\mu$, $\Lambda$, $\Pi$ and $\Phi$. In this section, we present constraints on the functional form of these quantities, which follow from elementary statistical relations and statistical isotropy. These constraints are useful to narrow down the possible functional forms of the conditional averages.\par
We start out with the pressure gradient. As already mentioned, its averaged projection on the velocity vector has to vanish because of incompressibility and homogeneity. This in turn gives an integral constraint on $\Pi(v)$\footnote{We omit the $t$-dependence to simplify the presentation. However, we do not assume stationarity unless explicitly mentioned.}:
\begin{equation}
  0 = \left \langle \bs{u}\cdot\bs{\nabla} p \right \rangle = \int_{0}^{\infty} \mathrm{d}v \, \left \langle \bs{u}\cdot\bs{\nabla} p  \big | v \right \rangle \tilde f(v) = -\int_{0}^{\infty} \mathrm{d}v \, v\,\Pi(v) \tilde f(v).
  \label{pressure-int-constraint}
\end{equation}
With an analogous argumentation one also finds constraints for $\Lambda$, $\lambda$ and $\mu$,
\begin{align}
 -\left\langle \varepsilon \right\rangle &= \left\langle \nu \bs u \cdot \Delta \bs u\right\rangle=  \int_{0}^{\infty} \mathrm{d}v \, v\,\Lambda(v) \tilde f(v) \label{Lambda-int-constraint}, \\
 \left\langle \varepsilon \right\rangle &= \left\langle \mathrm{Tr}(\mat{D}) \right\rangle=  \int_{0}^{\infty} \mathrm{d}v \, [\lambda(v)+2\mu(v)] \tilde f(v) \label{lambda-mu-int-constraint},
\end{align}
and in the stationary case also for the forcing
\begin{equation}
  \left\langle \varepsilon \right\rangle = \left\langle \bs u \cdot \bs F\right\rangle=  \int_{0}^{\infty} \mathrm{d}v \, v\,\Phi(v) \tilde f(v).
  \label{Phi-int-constraint}
\end{equation}

A second class of constraints can be deduced from statistical isotropy, which states that the conditional averages are invariant under rotations e.g.
\begin{equation}
  \left\langle \bs{\nabla} p \big| \mat{R}\bs v\right\rangle=\mat{R}\left\langle \bs{\nabla} p \big| \bs v\right\rangle,\qquad \mat{R}\in\mathrm{SO}(3).
\end{equation}
From this relation, it follows that the \emph{constant} vector $\left\langle \bs{\nabla} p \big| \bs v=\bs{0}\right\rangle$ is invariant under rotations. However, this is possible only for the zero vector, thus
\begin{align}
  \left\langle \bs{\nabla} p \big| \bs v=\bs{0}\right\rangle=\bs{0} \quad \Rightarrow & \quad  \Pi(0)=\lim_{v\rightarrow0}\hat{\bs v}\cdot\left\langle -\bs{\nabla} p \big| \bs v\right\rangle=0 .
\end{align}
Note that the limit $\lim_{v\rightarrow0}\hat{\bs v}$ is not defined; however, the boundedness of $\hat{\bs v}$ is enough for the above result. In summary, we find
\begin{align}
  \left\langle \bs{\nabla} p \big| \bs v=\bs{0}\right\rangle=\bs{0}           \quad \Rightarrow & \quad \Pi(0)=0 \label{pressure-zero-constraint}\\
  \left\langle \nu \Delta \bs u \big| \bs v=\bs{0}\right\rangle=\bs{0}    \quad \Rightarrow & \quad \Lambda(0)=0 \label{Lambda-zero-constraint}\\
  \left\langle \bs F \big| \bs v=\bs{0}\right\rangle=\bs{0}               \quad \Rightarrow & \quad \Phi(0)=0 \label{force-zero-constraint} .
\end{align}
This type of argument can also be applied to tensor-valued isotropic functions. In three dimensions the only constant isotropic tensor of rank two is $\delta_{ij}$, which leads to constraints on the eigenvalues
\begin{equation}
  D_{ij}(\bs v=\bs{0}) \sim \delta_{ij}\quad\Rightarrow\quad\lambda(0)=\mu(0).
  \label{lambda-eq-mu-at-zero}
\end{equation}
We can go even further and use this type of argument to obtain constraints on the derivatives at $\bs{v}=\bs 0$. For this, first note that the isotropic forms of vector- and tensor-valued functions \eqref{eq:isotropy_a} and \eqref{eq:isotropy_B} are also invariant under reflections. Therefore, all the isotropic functions considered here are actually invariant under transformations in $\mathrm{O}(3)$. This means that in our case, the assumption of invariance under $\mathrm{SO}(3)$ yields invariance under $\mathrm{O}(3)$. Now, it can be easily checked that if a tensor-valued function is invariant under some transformations, then so are also its derivatives. Therefore, the $n$-th derivative of $D(\bs{v})$ at $\bs{v}=\bs 0$ is a constant $(n+2)$-th-order tensor, invariant under $\mathrm{O}(3)$. For $n$ being odd, the derivative has to vanish, because the only constant tensor of odd order which is invariant under reflections is the zero tensor and it follows that
\begin{equation}
  \left. \frac{\partial}{\partial v_{i_1}}\cdots\frac{\partial}{\partial v_{i_n}} D_{ij}\right|_{\bs{v}=0}=0 
  \quad \Rightarrow \quad 
  \left.\frac{\mathrm{d}^n\lambda}{\mathrm{d}v^n}\right|_{v=0}=
  \left.\frac{\mathrm{d}^n\mu}{\mathrm{d}v^n}\right|_{v=0}=0, 
\end{equation}
where $n$ is odd.
With the very same argumentation applied to vectors (i.e. tensors of rank one) one also finds
\begin{align}
  \left.\frac{\mathrm{d}^n\Pi}{\mathrm{d}v^n}\right|_{v=0}=\left.\frac{\mathrm{d}^n\Lambda}{\mathrm{d}v^n}\right|_{v=0}=\left.\frac{\mathrm{d}^n\Phi}{\mathrm{d}v^n}\right|_{v=0}=0,
  \label{derivative-general}
\end{align}
where $n$ is even.

This shows that in isotropic turbulence, the series expansion of $\lambda$ and $\mu$ contains only even powers, whereas that of $\Pi$, $\Lambda$ and $\Phi$ contains only odd. These constraints have also to be kept in mind, when approximating the above functions with polynomials.

\subsection{A Simple Analytical Closure}\label{sec:simple-closure}

We now try to find the most simple functional forms of the conditional averages which still fulfil the constraints described in the previous section. By `most simple` we mean the lowest-order polynomial. For $\Lambda$, a constant is not allowed because of \eqref{Lambda-int-constraint} and \eqref{Lambda-zero-constraint}. We choose a linear function and determine the pre-factor through \eqref{Lambda-int-constraint}. We obtain
\begin{equation}
  \Lambda(v)=-\frac{\langle\varepsilon\rangle}{3\sigma^2}v
\end{equation}
where $\sigma=\sqrt{\langle \bs{u}^2\rangle/3}$ is the standard deviation of the velocity. In the stationary case, we further find $\Phi(v)=\langle\varepsilon\rangle/(3\sigma^2)v$. For $\lambda$ and $\mu$, the most simple choice which is consistent with \eqref{lambda-mu-int-constraint} and \eqref{lambda-eq-mu-at-zero} is a constant:
\begin{equation}
  \lambda(v)=\mu(v)=\frac{\langle\varepsilon\rangle}{3}.
\end{equation}
Inserting this ansatz for $\lambda$, $\mu$, $\Lambda$ into the homogeneous solution \eqref{eq:homosol} yields
\begin{equation}
  \tilde{f}(v;t)=\sqrt{\frac{2}{\pi}}\frac{v^2}{\sigma(t)^3}\exp\left( -\frac{1}{2} \frac{v^2}{\sigma(t)^2} \right),
  \label{gaussian-angleintegrated}
\end{equation}
which is the angle-integrated Gaussian distribution, i.e. the velocity \emph{vector} is distributed according to a Gaussian 
\begin{equation}
  f(\bs{v};t)=\frac{1}{(2\pi\sigma(t)^{2})^{3/2}}\exp\left( -\frac{1}{2}\frac{\bs{v}^2}{\sigma(t)^2} \right),
  \label{gaussian}
\end{equation}
with standard deviation $\sigma(t)$. This solution is valid for decaying as well as stationary turbulence, where $\sigma(t)$ is a monotonously decaying function in the first and a constant in the latter case. 
The above solution rests upon the homogeneity relation \eqref{eq:pdfhomoiso}, which in the stationary case is equivalent to the evolution equation \eqref{eq:pdfvelhomoiso}. However, for decaying turbulence, this evolution equation gives another possibility to obtain the PDF which we demonstrate in the following. For this, we choose the most simple functional form for the pressure gradient which is consistent with \eqref{pressure-int-constraint} and \eqref{pressure-zero-constraint}, namely $\Pi(v)=0$. This ansatz can be also viewed as an approximation which neglects the statistical dependence of the pressure gradient and the velocity. Furthermore, $\lambda$ and $\mu$ are chosen as above, representing statistical independence of the dissipation tensor and the velocity. With these assumptions, the evolution equation \eqref{eq:pdfvelhomo2} simplifies to
\begin{equation}\label{eq:simpleapprox}
  \frac{\partial}{\partial t}f=-\frac{\langle\varepsilon\rangle(t)}{3}\Delta_{\bs v}f.
\end{equation}
This is exactly the equation derived in \cite[]{hosokawa08pre} by neglecting pressure contributions and approximating the coupling of the dissipative term to the two-point PDF. By a change of variables \cite[]{hosokawa08pre},
\begin{equation}
  \tau(t)=\frac{1}{3}E_{kin}(t)=\frac{1}{3}\left[ E_{kin}(t_0) - \int^t_{t_0} \mathrm{d}t' \, \langle \varepsilon \rangle(t') \right]=\frac{1}{2}\sigma(t)^2
  \label{tau-energy-sigma}
\end{equation}
the equation takes the form of the heat equation
\begin{equation}
  \frac{\partial}{\partial \tau}f=\Delta_{\bs v}f.
\end{equation}
The change of sign indicates that the process runs backwards in time. As stated before, when the fluid has come to rest, we find $f(\bs v)=\delta(\bs v)$, which now may serve as an initial condition. The solution reads as
\begin{equation}
  f(\bs v;t)=\frac{1}{(4\pi\tau(t))^{3/2}}\exp\left( -\frac{\bs v^2}{4\tau(t)} \right) ,
\end{equation}
i.e. we have a Gaussian PDF, whose time evolution is solely determined by the mean rate of energy dissipation. Note that this solution is consistent with \eqref{gaussian} as $\tau(t)=\sigma(t)^2/2$. Equation \eqref{tau-energy-sigma} highlights the connection between $\langle\varepsilon\rangle(t)$ and $\sigma(t)$ for decaying turbulence; because the energy dissipation reduces the kinetic energy of the flow, the PDF becomes narrower and narrower as a function of time.\par
The results presented in this section should not be understood as a justification for the Gaussianity of the velocity as they have been obtained by using purely mathematical arguments, which do not account for any physical properties of the velocity field. For example, the argumentation using the homogeneous solution (presented at the beginning of this section) can also be applied to the vorticity field, which is known to exhibit a highly non-Gaussian distribution. The validity of the above results depends on how well the simple choices for the conditional averages agree with their functional forms measured in turbulence. For this comparison, we refer to the next sections, where the numerical results will be presented.\par
With the results of this section, the Gaussian distribution can be viewed as the most simple solution of the equations describing the velocity PDF, which is consistent with the constraints presented above. In a sense it is a zeroth-order approximation to the real velocity PDF. Deviations of the conditional averages from their most simple forms lead to deviations from the Gaussian distribution. In the following sections, we will see that this zeroth-order approximation roughly describes the velocity PDF, but completely fails for the vorticity PDF.

\section{DNS results}
\subsection{Some comments on the numerical simulations}

\begin{table}
\begin{center}
  \begin{tabular}{rcccccccccc}
    \hline 
    & $N^3$ & $R_{\lambda}$ & $L$ & $T$ & $u_{rms}$ & $\nu$ & $\left\langle \varepsilon \right\rangle$ & $ k_{max}\eta$\\
    \hline 
    & $512^3$  &  112 & 1.55 & 2.86 & 0.543 & $10^{-3} $ & 0.103 & 2.03\\
    \hline
  \end{tabular}
\end{center}
\caption{Major simulation parameters. Number of grid points $N^3$, Reynolds number based on the Taylor micro-scale $R_{\lambda}$, integral length scale $L$, large-eddy turnover time $T$, root-mean-square velocity $u_{rms}$, kinematic viscosity $\nu$, average rate of energy dissipation $\left\langle \varepsilon \right\rangle$. Here $k_{max}\eta$ characterizes the spatial resolution of the smallest scales, where $\eta$ is the Kolmogorov length scale and $k_{max}=0.8 N/2$ is the highest resolved wavenumber (the factor 0.8 is due to the aliasing filter).}
\label{tab:simpara}
\end{table}
Before presenting the numerical results, some comments on the numerical scheme are in order. The presented results are generated with a standard, dealiased Fourier-pseudo-spectral code \cite[]{canuto87book,hou2007jcp} for the vorticity equation. The integration domain is a triply periodic box of box length $2\pi$. The time-stepping scheme is a third-order Runge--Kutta scheme \cite[]{shu88jcp}. For the statistically stationary simulations, a large-scale forcing has to be applied to the flow. Here, care has to be taken in order to fulfil the statistical symmetries we make use of in our theoretical framework. After numerous tests, we chose a large-scale forcing which conserves the energy of the flow. This forcing has been found to  deliver accurate results concerning the statistical symmetries. However, the results presented here do not depend on the exact form of the forcing, which has been verified using other forcing methods (e.g. the common method which holds the amplitudes of the Fourier coefficients in a band constant).\par
The numerical results presented in the following are obtained from a simulation whose simulation parameters are listed in table \ref{tab:simpara}. Care has been taken to produce a well-resolved simulation with an integral length scale which is not too large compared with the box length of $2\pi$. We have found that the statistical symmetries, especially isotropy, are not valid for a snapshot of the velocity field at a single point in time, but can be obtained through time- or ensemble-averaging, respectively. This is due to long-range correlation of the velocity and is not as pronounced for the vorticity, a short-range correlated quantity. To ensure the validity of the statistical symmetries, the averages for the stationary case have been obtained through spatial and temporal averaging over more than 150 large-eddy turnover times, whereas for decaying turbulence a spatial and an ensemble average over 12 realizations has been applied.\par
It may be doubted that the results obtained from a single numerical set-up are of universal nature, for example there may be influences of the imposed (periodic) boundary conditions. This critique, of course, also applies to any experimental result, so that a compilation of data from many different numerical and experimental set-ups would be useful to discuss the issue on a more general level. However, the results presented in the following will make a good point that there are deviations from the Gaussian shape in both stationary and decaying turbulence.
\par

\subsection{Stationary Turbulence}\label{sec:stationary-turbulence}

The homogeneous and stationary solutions \eqref{eq:homosol} and \eqref{eq:statsol} of the PDF equations allow a detailed analysis of the connection between the functional shape of the PDF and the statistical correlations of the different dynamical quantities, represented by the conditional averages $\Pi$, $\Lambda$, $\Phi$, $\lambda$ and $\mu$. This analysis is presented in this section for the case of stationary turbulence.

\begin{figure}
\centering\includegraphics[width=0.65\textwidth]{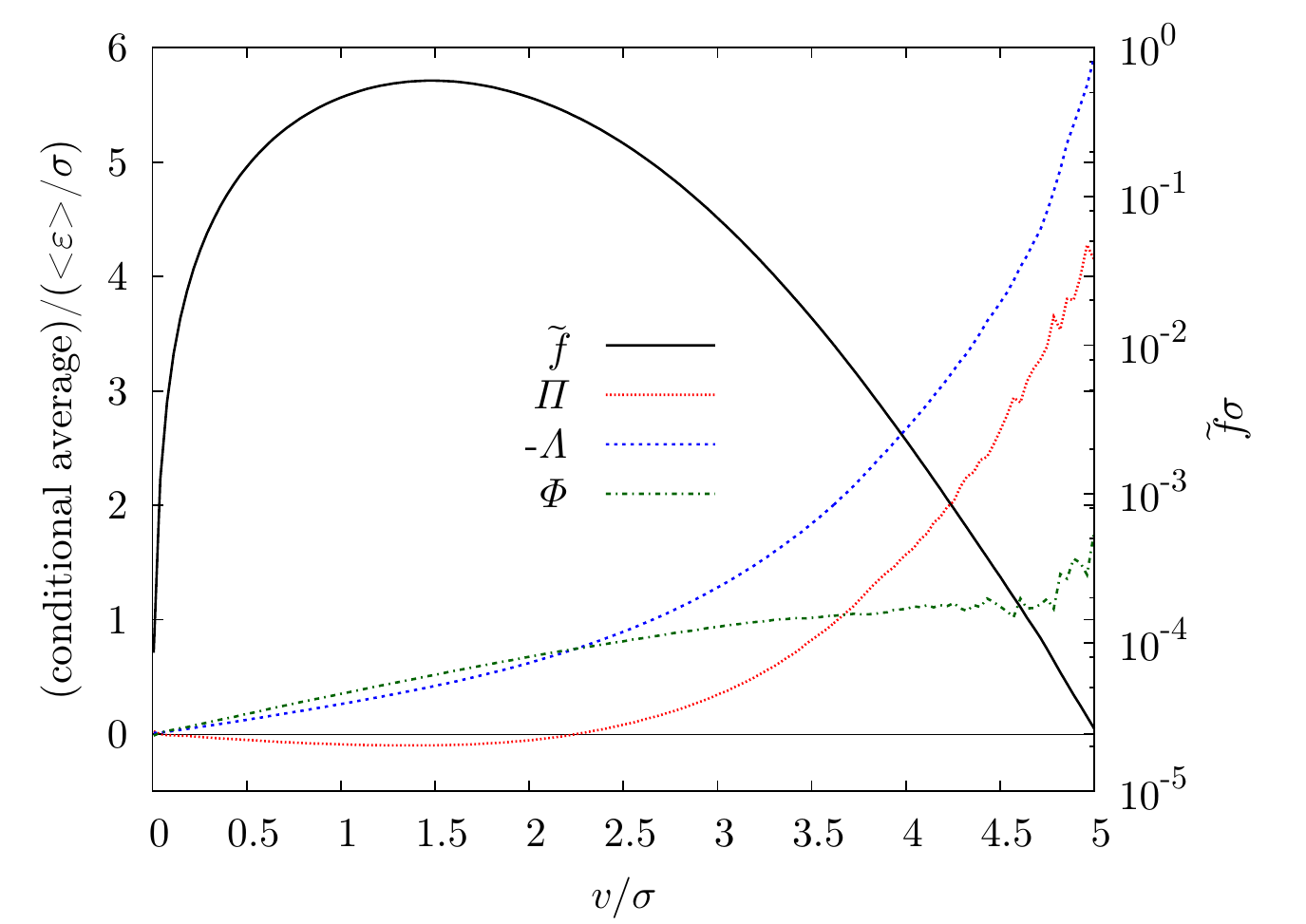}
\caption{The conditional averages $\Pi(v)$, $\Lambda(v)$, $\Phi(v)$ and the PDF $\tilde f(v)$. The $v$-axis is scaled with the standard deviation of the velocity $\sigma=u_{rms}=\sqrt{\left\langle \bs u^2/3\right\rangle}$. The conditional averages show an explicit $v$-dependence indicating statistical correlations.}
\label{fig:PILAPHI}
\end{figure} 
To start with, we consider the diffusive term, shown in figure \ref{fig:PILAPHI}. It exhibits a negative correlation with the velocity, indicating that a fluid element will on average be decelerated due to viscous forces. The pressure term, however, displays an interesting zero-crossing. This means that a fluid particle will on average be decelerated for low values of velocity and accelerated for high values of velocity. If the pressure term is non-vanishing, it \emph{has} to exhibit this zero-crossing, because the integral constraint \eqref{pressure-int-constraint} can be fulfilled only through a zero-crossing or vanishing of $\Pi(v)$. Our numerical investigations show that the pressure contributions to the single-point velocity PDF cannot be neglected, in contrast to the assumptions of some recent theories \cite[]{tatsumi04fdr,hosokawa08pre}. This is especially true for high velocities, where we see a strong $v$-dependence of $\Pi$ and hence a clear statistical dependence of the pressure gradient and the absolute value of the velocity.\par
The conditionally averaged forcing term in figure \ref{fig:PILAPHI} was computed based on the conditional balance \eqref{eq:probcurrent}. This is because the forcing in our numerical scheme is implicitly applied to the vorticity field and is not available as an additive field that we can average conditionally. However, some careful tests have been performed to ensure that this balance actually holds. An interesting feature of $\Phi$ is that it seems to saturate for large values of $v$ in contrast to $\Pi$ and $\Lambda$, suggesting that the forcing plays a minor role for the statistics of high velocities. In \cite[]{falkovich97prl} it has been argued that the tails of the velocity PDF are related to those of the forcing. Our numerical results, however, suggest that the main contribution comes from the dynamical effects of the pressure gradient and the velocity diffusion. Furthermore, the saturation of $\Phi$ suggests a simple relation between $\Pi$ and $\Lambda$ for high velocities, namely an approximately constant offset. This shows that for high velocities, the statistical dependences of the pressure gradient and the velocity diffusion on the velocity are strongly related. A natural question concerning the conditionally averaged forcing is whether its form depends on the type of forcing. We have tested two other common forcing methods (these methods manipulate the vorticity in a Fourier band by holding either the Fourier coefficients or their amplitudes constant), finding that the general shape of $\Phi$ does not depend on the forcing method.\par

\begin{figure}
  \centering\includegraphics[width=0.65\textwidth]{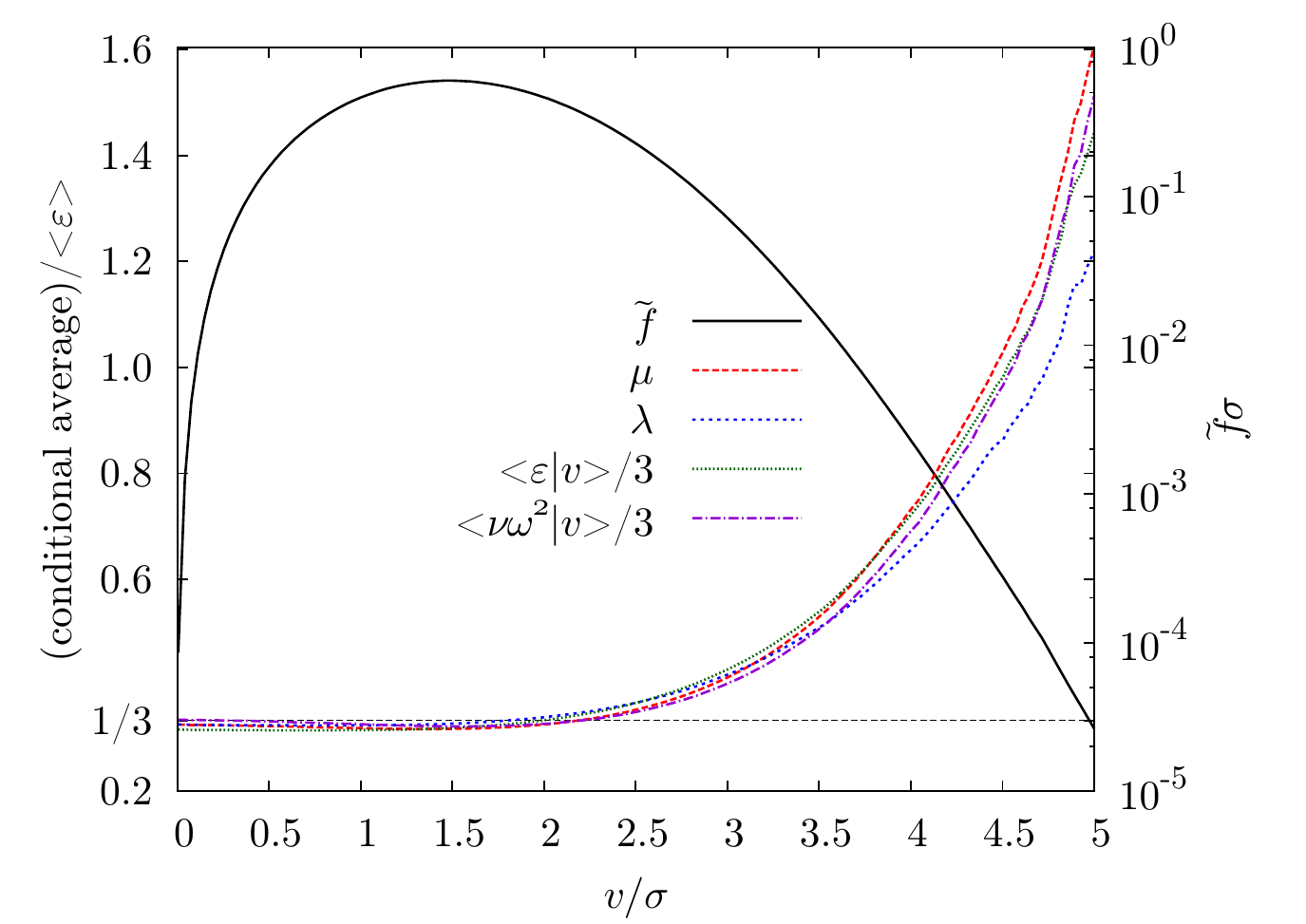}
  \caption{The conditional averages $\mu(v)$, $\lambda(v)$, $\left\langle \varepsilon |v \right\rangle$, $\left\langle \nu \omega^2 |v \right\rangle$ and the PDF $\tilde f(v)$. The conditional averages show an explicit $v$-dependence indicating statistical correlations.}
  \label{fig:mulaepsz}
\end{figure}
Now let us proceed to investigate the conditional energy dissipation tensor, its eigenvalues are shown in figure {\ref{fig:mulaepsz}} together with the PDF $\tilde f(v)$. Both eigenvalues have a similar dependence on $v$, they are almost constant for small and medium velocities, but display a strong $v$-dependence for high velocities. This shows that the occurrence of high velocities is statistically correlated with strong dissipation and enstrophy.\par
\begin{figure}
  \centering\includegraphics[width=0.65\textwidth]{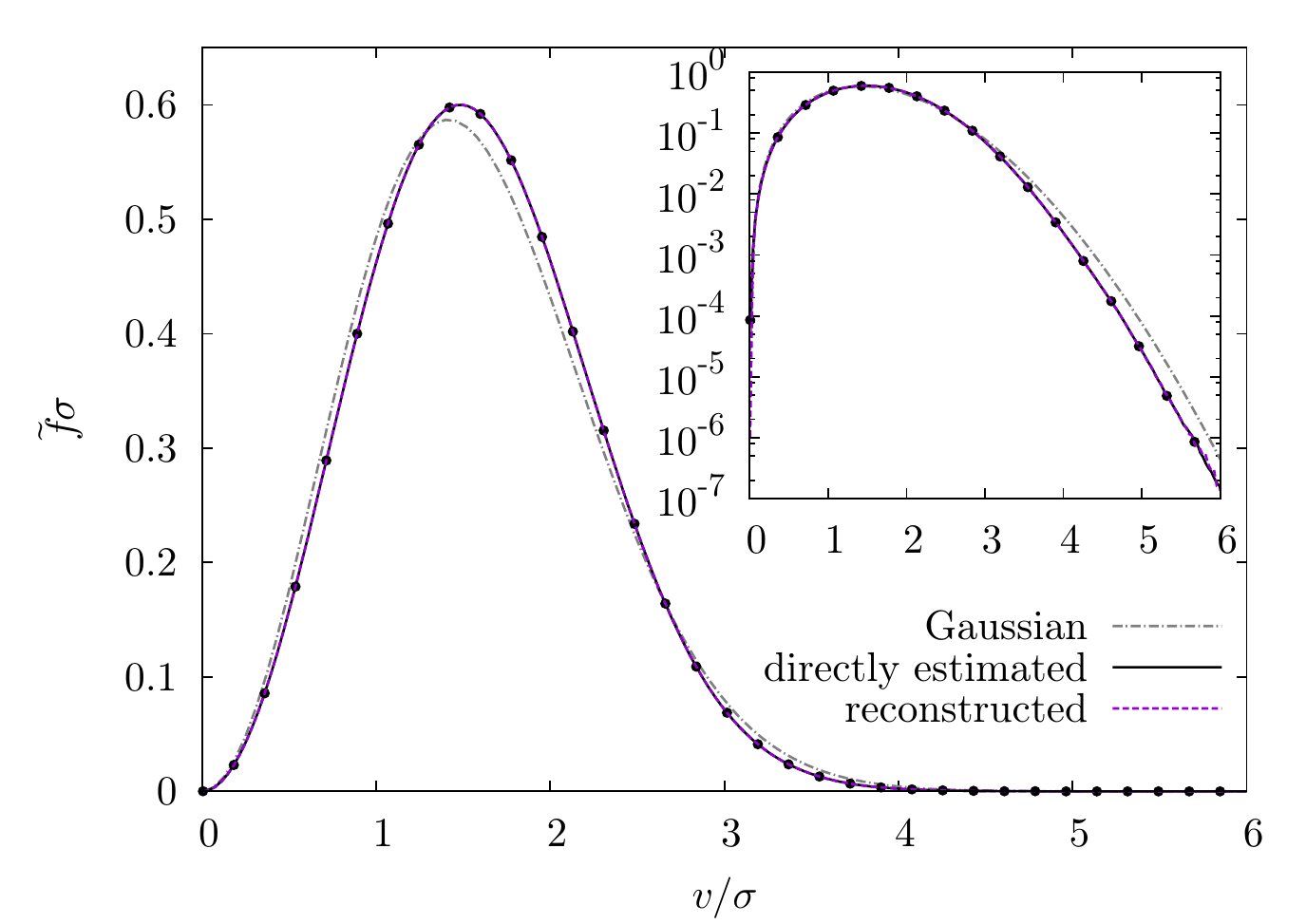}
  \caption{Comparison of the PDF $\tilde f(v)$ with the evaluation of \eqref{eq:homosol}. The reconstructed PDF collapses with the directly estimated PDF; therefore, the latter is marked by black dots to indicate its presence. The excellent agreement demonstrates the consistency of the theoretical results. A comparison with an angle-integrated Gaussian shows that, although being close to Gaussian, significant deviations exist.}
  \label{fig:recpdf}
\end{figure}
The conditional averages shown in figures \ref{fig:PILAPHI} and \ref{fig:mulaepsz} have been tested to obey the constraints presented in \S\,\ref{sec:Constraints}. The validity of the constraints at $v=0$ can be seen in the figures, whereas the integral constraints have been verified numerically. As a test for consistency, the conditional averages are inserted into \eqref{eq:homosol} to calculate the homogeneous PDF, and the result is presented in figure \ref{fig:recpdf}. An excellent agreement with the PDF directly estimated from the numerical data is found. The same results can be obtained using \eqref{eq:statsol} as it is equivalent to \eqref{eq:homosol} in the stationary case. A comparison with an angle-integrated Gaussian \eqref{gaussian-angleintegrated} shows that, although the PDF is similar to Gaussian, significant deviations occur. Especially notable is the sub-Gaussian tail, which has also been found in experiments and other numerical simulations \cite[]{vincent91jfm,noullez97jfm,gotoh02pof}.\par
It is well known for a long time that the velocity PDF is similar to a Gaussian distribution. However, this fact (coming from experiments and simulations) is \textit{a priori} not clear at all. One can certainly arrive at a Gaussian distribution by neglecting certain terms  or assuming the most simple forms for the conditional averages (see \S\,\ref{sec:simple-closure}). However, as can be clearly seen in figures \ref{fig:PILAPHI} and \ref{fig:mulaepsz}, those assumptions are violated. Furthermore, some of the arguments can also be applied to the vorticity which is known to exhibit a highly non-Gaussian PDF. This shows that one has to be very careful with the utilization of such simple arguments. Therefore, here we rather use the measured conditional averages to understand the similarity of the velocity PDF to the Gaussian distribution.
The question sought after is, how it is possible to find strong statistical correlations for the various conditional averages and the velocity, but still only moderate deviations from Gaussianity for the PDF.\par
 As already stated, there are clear deviations of the conditional averages from the simple forms suggested in \S\,\ref{sec:simple-closure}. We see that for small and medium velocities, the assumption of linear $\Lambda$ (or equivalently of linear $\Phi$ and vanishing $\Pi$) and constant $\lambda,\mu$ can be seen as a rough approximation. Because of this, the PDF stays close to a Gaussian in this range of velocities. However, for high velocities we see a strong $v$-dependence of $\lambda$, $\mu$ and a functional form of $\Lambda$ which is clearly not linear. Still, only moderate deviations from a Gaussian tail occur. To understand this, let us examine again the relation between the PDF and the conditional averages:
\begin{equation}
  \tilde f(v)=\frac{{\cal N}}{\lambda(v)} \exp \int_{v_0}^v \mathrm{d}v' \, \frac{ \Lambda(v')+\frac{2}{v'}\mu(v')}{\lambda(v')}.
  \label{eq:homsol-again}
\end{equation}
We see that only the quotients $\Lambda/\lambda$ and $\mu/\lambda$ enter the exponential function. As $-\Lambda$ and $\lambda$ both are bended upwards for high velocities, their quotient might still be linear, i.e. the deviations from the simple forms of \S\,\ref{sec:simple-closure} might compensate each other. Indeed, this is roughly true as can be seen in figure \ref{fig:integrands-analyzed}, which also shows that $\lambda$ and $\mu$ are roughly equal. Furthermore, a comparison with the case of vorticity is shown, which will be discussed later. Assuming for the moment a linear $\Lambda/\lambda$ and $\mu/\lambda=1$ yields a Gaussian shape for the exponential factor in \eqref{eq:homsol-again}. In this case, the second factor, $1/\lambda$, is then responsible for (slight) modifications of the Gaussian shape. Therefore, it appears that the compensation of $\Lambda$ and $\lambda$ to yield an approximately linear quotient $\Lambda/\lambda$ is crucial for the similarity of the velocity PDF to the Gaussian distribution. For example, if one artificially sets $\lambda=\mu=\langle\varepsilon\rangle/3$ as in \S\,\ref{sec:simple-closure}, but still uses the measured and highly nonlinear $\Lambda$ for the reconstruction, one obtains a PDF which decays much faster than a Gaussian.\par
To discuss the issue of the sub-Gaussian tails, we need to take into account the deviations of the quotients from a linear and constant function. A detailed analysis shows that the deviations from Gaussianity cannot be solely attributed to the pre-factor $1/\lambda$, but are also due to the aforementioned deviations of the quotients. What makes the issue even more complicated is that the relative amount of contribution to the deviations by the two factors in \eqref{eq:homsol-again} seems to vary with the Reynolds number, as has been checked with a number of simulations not presented here. This shows that the detailed properties of the PDF of the velocity are due to a subtle interplay of the statistical dependences of dynamical quantities such as the dissipation, the velocity diffusion or the pressure gradient.

\subsubsection{Approximating the Dissipation Tensor}
In view of the complicated statistical dependences, it is natural to ask for approximations which simplify the matter but still describe the velocity PDF reasonably well e.g. maintain the sub-Gaussianity of the velocity PDF. It turns out that the conditional dissipation tensor allows simplifications which contain some interesting physical insights. It appears from figure \ref{fig:mulaepsz} that the eigenvalues of $\mat{D}$ have a similar functional form, suggesting the approximation $\lambda(v)\approx\mu(v)$. Thus, all eigenvalues of $\mat{D}$ are approximately equal, which basically states that the conditional dissipation tensor does not contain any directional information and is proportional to the identity matrix
\begin{equation}
  D_{ij}(\bs v)\approx\frac{1}{3}\mathrm{Tr}(\mat{D})\,\delta_{ij}=\frac{1}{6}\langle \varepsilon + \nu \omega^2 |v \rangle \, \delta_{ij}.
\end{equation}
In this approximation, $\mat{D}$ depends only on the magnitude of the velocity, i.e. the dissipation tensor $\nu\mat{AA}^{\mathrm T}$ and the \emph{direction} of the velocity vector are uncorrelated. Note that this property is not a consequence of the statistical isotropy, but rather a special property of the velocity in turbulent flows (this will become more clear when we examine the dissipation tensor of the vorticity).\par
The equality of the eigenvalues may be motivated with a simple scale-separation argument. The velocity itself is known to be a rather smooth vector field varying quite slowly in space. Compared to that, the gradients vary on a much shorter scale (see figure \ref{fig:vortvel}). The numerical results suggest that quantities like, for example, the dissipation depend on the absolute value $v$. If we now assume that the dependence on $v$ is crucial, but due to scale separation no correlation between the direction of $\bs u$ and the velocity gradient tensor exists, we may consider a projection on $\hat{\bs u}$ as a projection on a random direction. This immediately gives
\begin{equation}
  \langle \nu (\mat{A}^{\mathrm T}\hat{\bs u})^2 | v \rangle=\langle \nu \hat{\bs u}\mat{AA}^{\mathrm T}\hat{\bs u} | v \rangle=\frac{1}{6}\langle \varepsilon + \nu \omega^2 | v \rangle,
\end{equation}
i.e. averaging the projection of a tensor on a random direction yields one-third of the conditionally averaged trace of the tensor. This in turn yields $\lambda(v)=\mu(v)$, which is suggested by the numerical results as an approximation.\par
As the trace is the central quantity, one may wonder if further simplifications are possible. To this end, we consider the average proportionality of enstrophy and dissipation in homogeneous turbulence, $\langle \varepsilon \rangle=\langle \nu \omega^2 \rangle $, which comes due to $\langle \Delta p \rangle=0$. It is now straightforward to ask whether this balance also holds in a stronger sense, when the conditional average is considered. The numerical evaluation of these terms is also presented in figure \ref{fig:mulaepsz}, supporting that this stronger equality actually holds. As a consequence, the approximation $\frac{1}{2}\langle \varepsilon + \nu \omega^2 | v \rangle \approx \langle \varepsilon | v \rangle$ seems reasonable. Taking this all together, the conditional dissipation tensor can be approximated as
\begin{equation}
  D_{ij}(\bs v) \approx \frac{1}{3} \langle \varepsilon | v \rangle \, \delta_{ij},
\end{equation}
depending only on the conditionally averaged rate of kinetic energy dissipation.\par
It may be hypothesized that the approximate relations so far only hold asymptotically with e.g. increasing Reynolds number. By comparing simulations with different Reynolds numbers (up to $Re_\lambda\approx200$), we were not able to clearly verify (or falsify) this statement; hence, simulations with significantly higher Reynolds numbers are needed to answer this question.\par
In the light of these simplifications, it is interesting to reconsider the functional form of the velocity PDF. Equations \eqref{eq:isopdf} and \eqref{eq:homsol-again} together with $\lambda(v)\approx\mu(v)\approx \langle \varepsilon | v \rangle/3$ instantly yield
\begin{align}\label{eq:simplepdf}
  f(\bs v)&=\frac{{\cal N}}{\langle \varepsilon | v \rangle} \exp \int_{v_0}^v \mathrm{d} v' \, 3 \, \frac{\langle \nu \hat{\bs u}\cdot \Delta \bs u | v' \rangle}{\langle \varepsilon | v' \rangle} \nonumber \\
  &=\frac{{\cal N}}{\langle \varepsilon | v \rangle} \exp \int_{v_0}^v \mathrm{d} v' \, 3 \, \frac{\langle \hat{\bs u}\cdot \bs{\nabla} p-\hat{\bs u}\cdot \bs{F} | v' \rangle}{\langle \varepsilon | v' \rangle} ,
\end{align}
indicating that the shape of the velocity PDF may be solely determined by the velocity diffusion and the rate of energy dissipation. As the diffusive term balances the sum of the pressure and the forcing term, the PDF is equivalently determined by how energy is injected into the system, redistributed by non-local pressure effects and finally dissipated.
\begin{figure}
  \centering\includegraphics[width=0.65\textwidth]{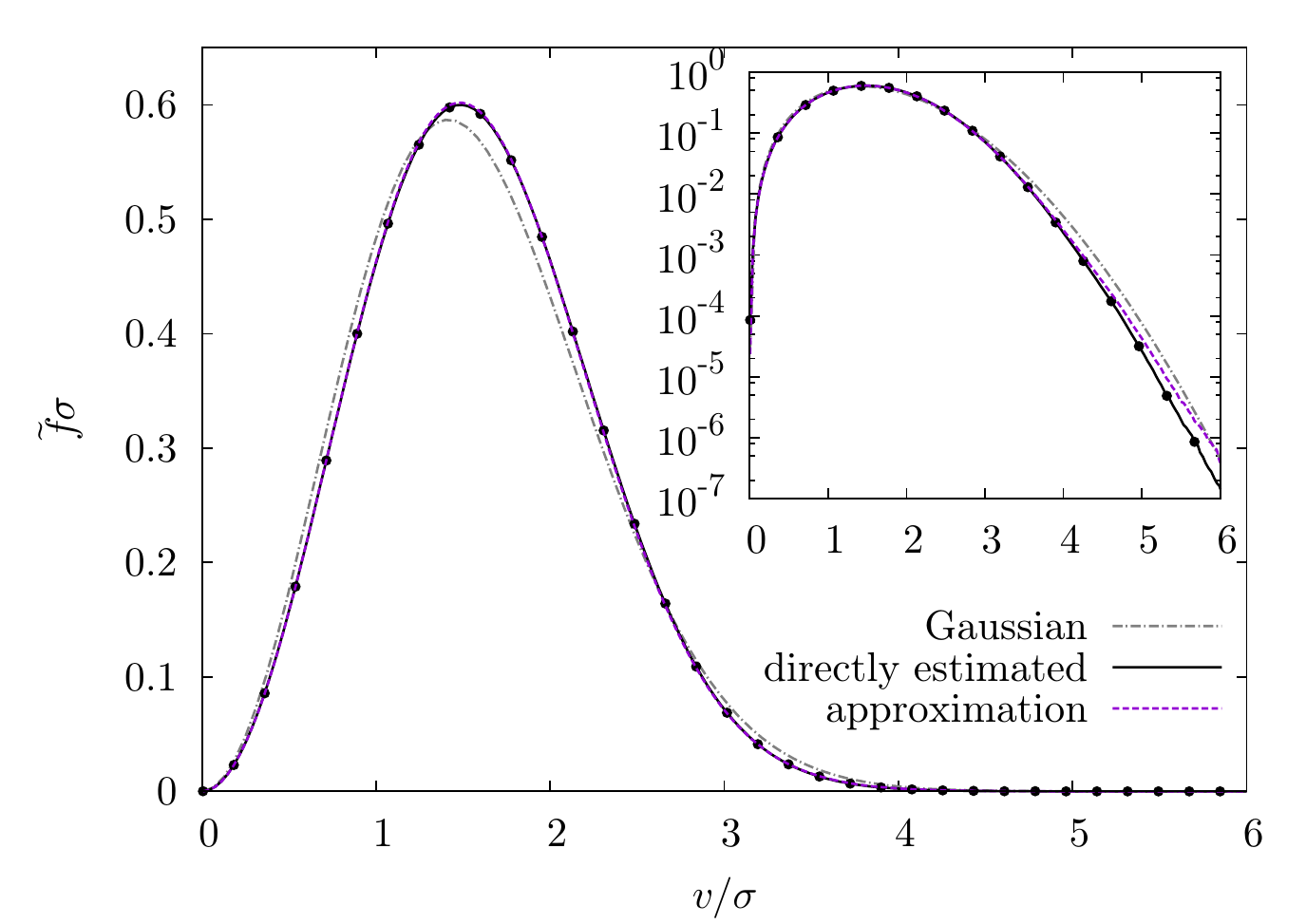}
  \caption{Comparison of the PDF $\tilde f(v)$ with the angle-integrated PDF of the approximation \eqref{eq:simplepdf}.}
  \label{fig:approximations}
\end{figure}
Figure \ref{fig:approximations} compares this approximation with the measured PDF and a Gaussian. We see that the above formula for the velocity PDF holds quite well for the core of the PDF, but leads to deviations in the tail. This (in addition to figure \ref{fig:mulaepsz}) shows that the assumption of absent correlation between the dissipation tensor and the direction of the velocity becomes less valid for high velocities. Nevertheless, this approximation describes the velocity significantly better than the Gaussian distribution.\par
We note here that the closeness of the velocity PDF to a Gaussian can be attributed to the approximate validity of
\begin{equation}
  \frac{\langle \nu \hat{\bs u} \cdot \Delta \bs u | v \rangle}{\langle \varepsilon | v \rangle} \approx-\frac{v}{3\sigma^2}.
  \label{eq:linear-quotient-approx}
\end{equation}
It describes the relation of the dissipation and the velocity diffusion, which is actually the dynamical effect responsible for the dissipation. It is the theoretical framework presented in the previous sections and in particular the formula \eqref{eq:homsol-again} which allows to relate the observation of closely Gaussian velocity PDF to the non-obvious relation above.

\subsection{Decaying Turbulence}

\begin{figure}
  \centering\includegraphics[width=0.65\textwidth]{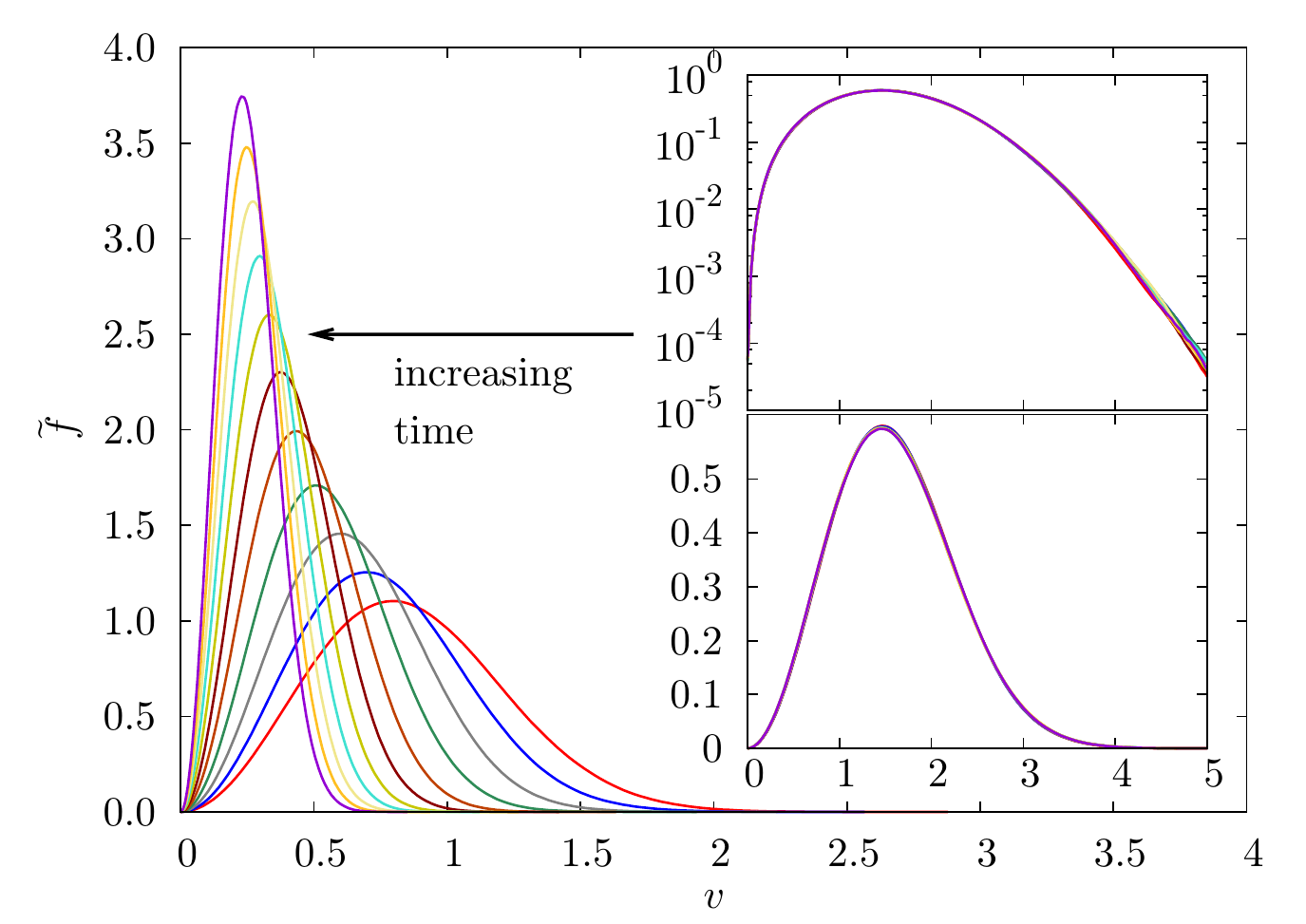}
  \caption{Velocity PDF $\tilde f(v;t)$ for decaying turbulence at time instants separated by $0.35T$, $T$ being the large-eddy turnover time of the initial condition. The insets show the PDFs normalized to unit variance, indicating the existence of a self-similar regime.}
  \label{fig:velpdftemp}
\end{figure}

\begin{figure}
  \centering\includegraphics[width=0.5\textwidth]{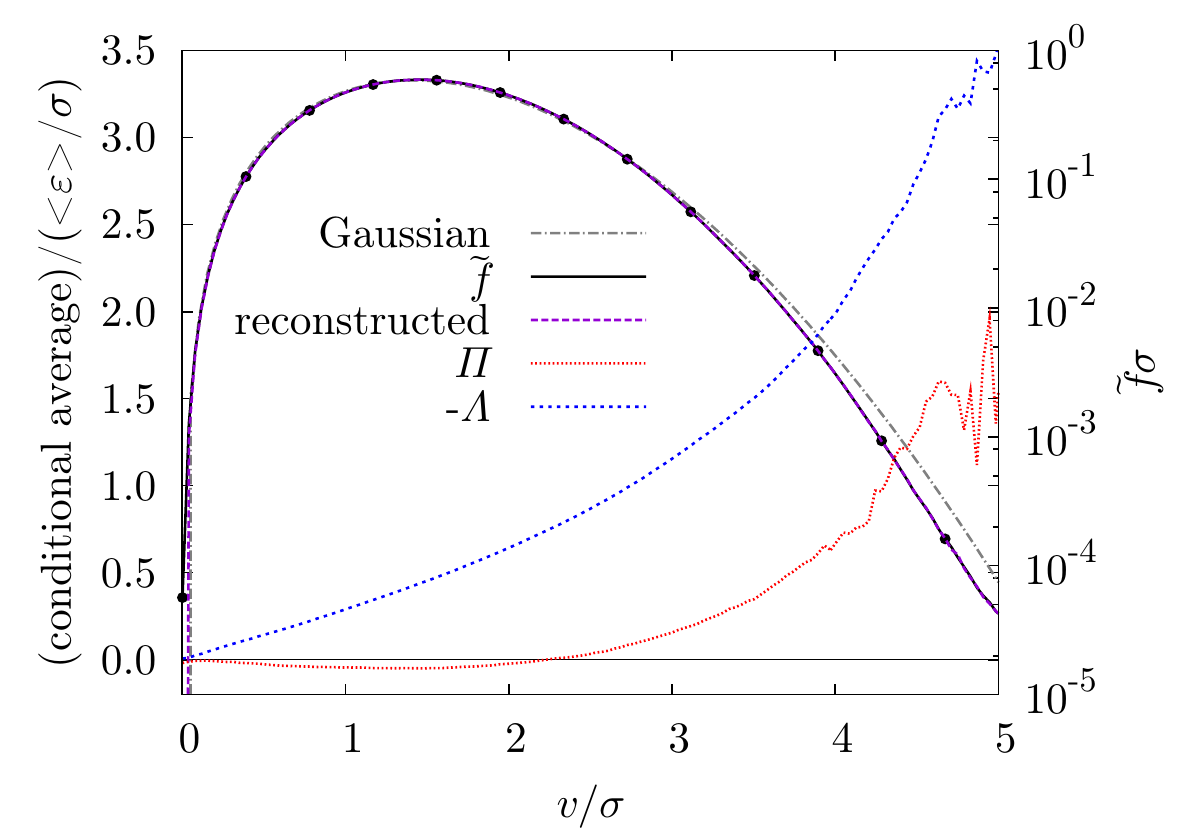}\includegraphics[width=0.5\textwidth]{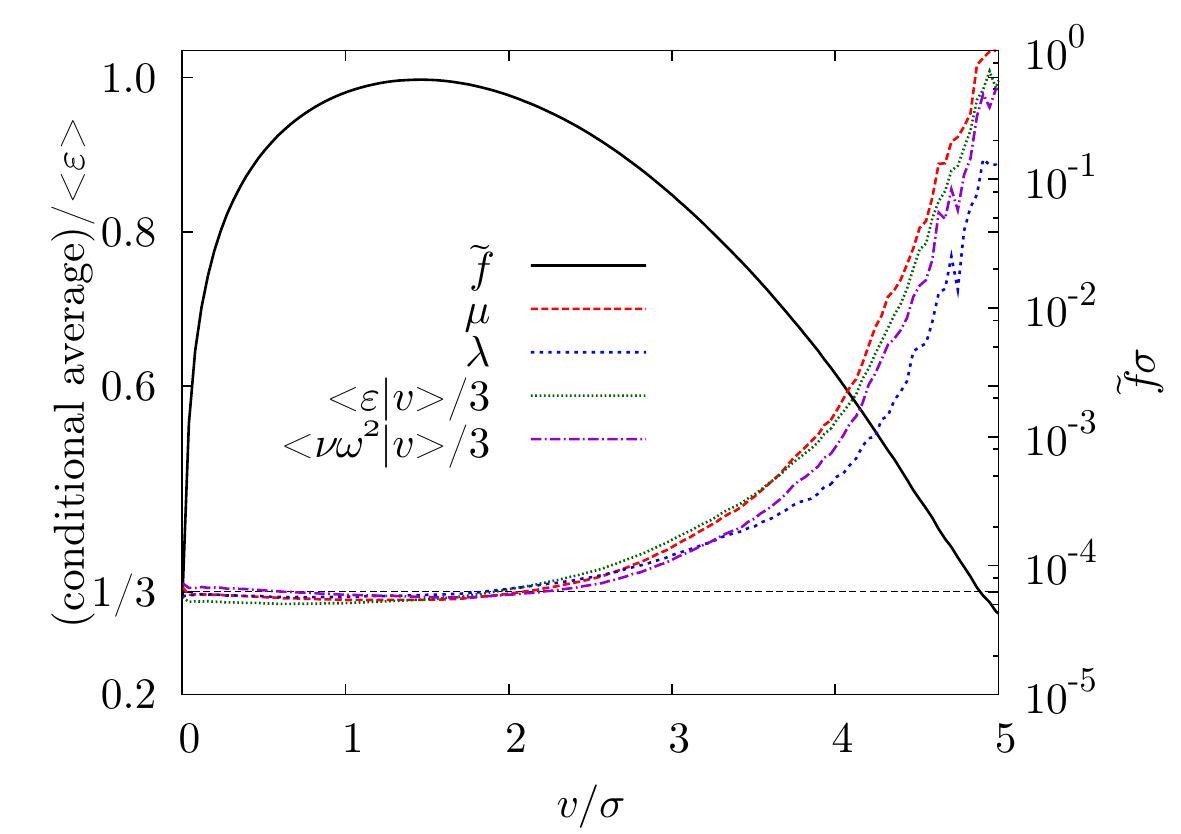}
  \caption{Comparison of $\tilde f(v)$ (directly estimated and reconstructed from \eqref{eq:homosol}) together with an angle-integrated Gaussian. A deviation from the Gaussian shape is also found in the decaying case. The structure of the conditional averages is similar to the case of stationary turbulence. The averages and the PDFs are computed from an ensemble for a decay time of $1.4T$.}
  \label{fig:recpdfdecay}
\end{figure}

\begin{figure}
  \centering\includegraphics[width=0.65\textwidth]{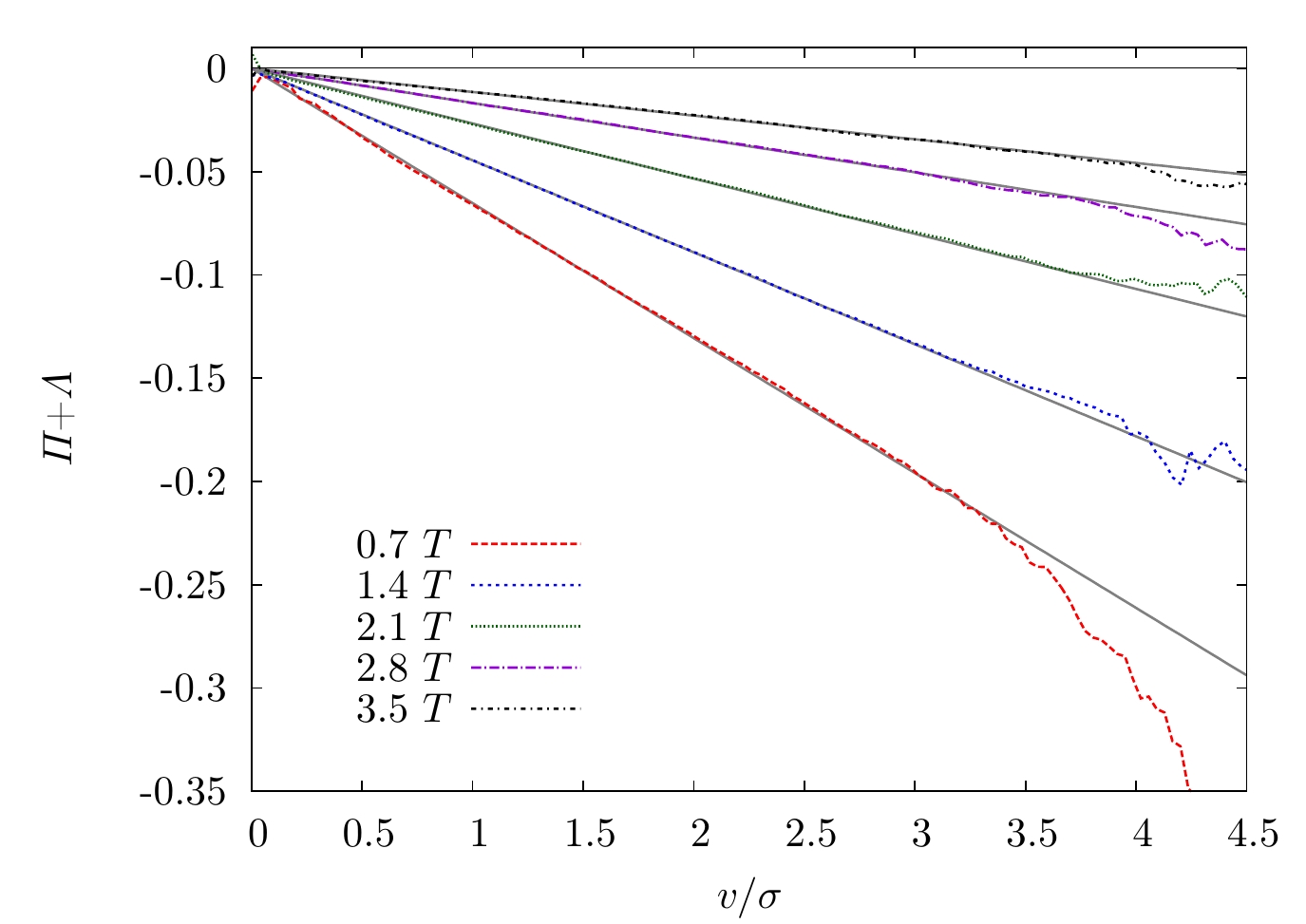}
  \caption{Sum of the pressure and diffusive terms for decaying turbulence together with the analytical relation \eqref{eq:selfsim} in straight lines. Slight deviations are visible after $0.7T$, which are negligible later on.}
  \label{fig:selfsim}
\end{figure}

\begin{figure}
  \centering\includegraphics[width=0.65\textwidth]{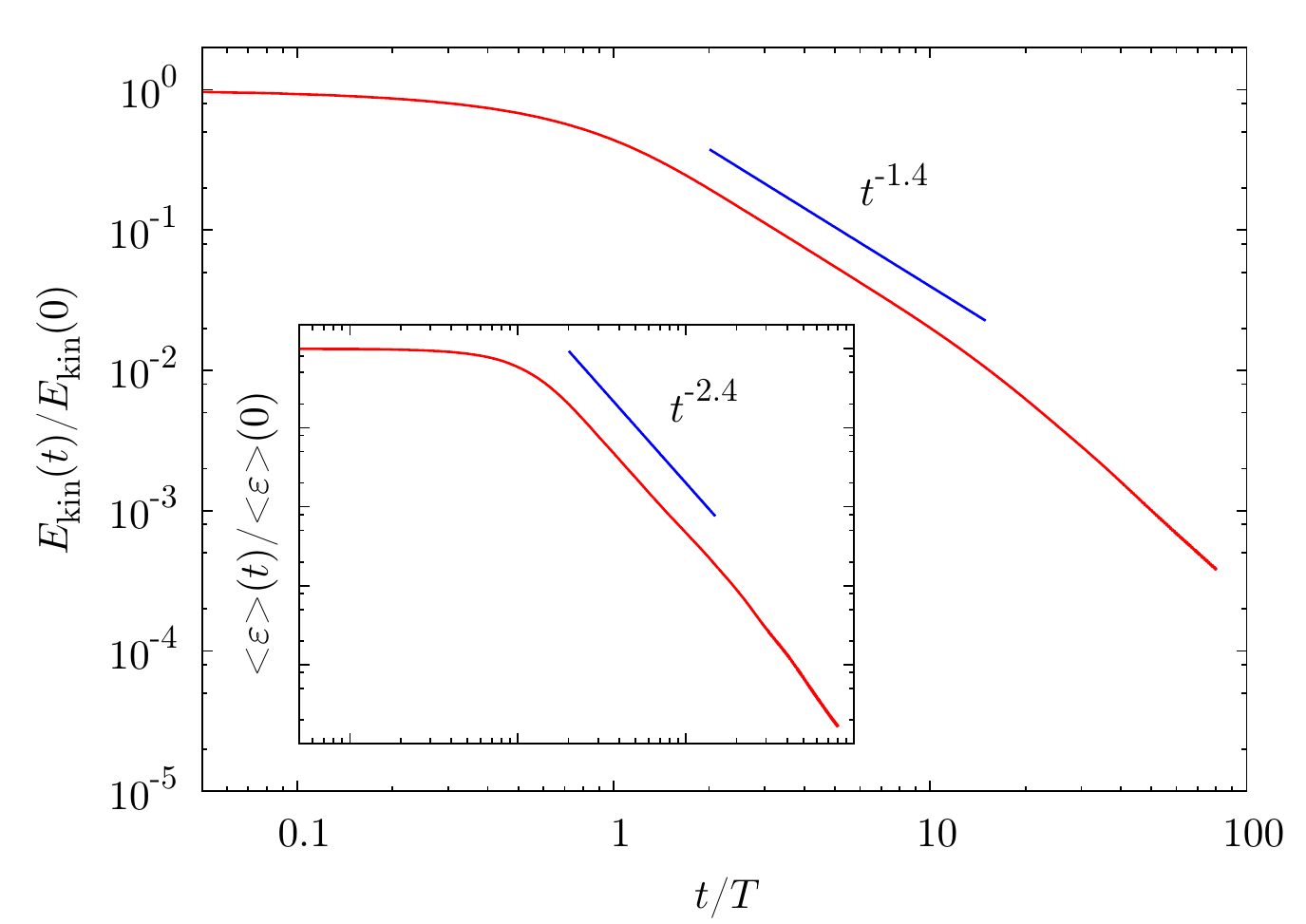}
  \caption{Kinetic energy for decaying turbulence as a function of time. After a transient the energy decays algebraically with an exponent close to $1.4$. The inset shows the rate of energy dissipation, which also exhibits an approximately algebraic decay with an exponent close to $2.4$.}
  \label{fig:decayenergy}
\end{figure}

We now study the shape and evolution of the velocity PDF in the case of decaying turbulence. The motivation for this is two fold. First, it has recently been hypothesized by \cite{hosokawa08pre} that decaying turbulence might be profoundly different from forced, stationary turbulence. As mentioned in \S\,\ref{sec:simple-closure}, Hosokawa found Gaussian solutions for decaying turbulence by approximate closure assumptions. Second, in the work by \cite{falkovich97prl}, the argument for sub-Gaussian tails of the velocity PDF relies strongly on the statistics of the external forcing. This effect, of course, is absent for decaying turbulence.\par
Figure \ref{fig:velpdftemp} shows the velocity PDF for decaying turbulence $\tilde f(v;t)$ for different points in time. The results were obtained by averaging over an ensemble of 12 independent simulations to increase the statistical quality. The variance of the PDFs decreases as a function of time as the kinetic energy of the system is dissipated. However, when rescaled to unit variance, the PDFs collapse, indicating a self-similar regime. Hence, once the PDF is specified at the beginning of this regime, it is determined subsequently according to
\begin{equation}\label{eq:selfsimilarpdf}
  \tilde f(v;t)= \frac{\sigma(t_0)}{\sigma(t)} \tilde f \left( \frac{\sigma(t_0)}{\sigma(t)} v; t_0 \right),
\end{equation}
where $\sigma(t)$ may be related to the kinetic energy or the rate of energy dissipation according to \eqref{tau-energy-sigma}. Turning to a closer investigation of the functional form of the PDF in this regime still reveals deviations from Gaussianity, as can be seen in figure \ref{fig:recpdfdecay}. Here, also the conditional averages of the pressure term, the diffusive term and the terms related to the conditional dissipation tensor are shown. They all show a functional form coinciding with the observations in stationary turbulence, indicating that the statistical correlations and the shape of the PDF do not fundamentally differ for decaying turbulence. As a check for consistency, the PDF is computed from the conditional averages by \eqref{eq:homosol}, which again performs very well. On the basis of these results, we can conclude that sub-Gaussian velocity PDFs are also found for decaying turbulence, which can be tracked down to the interplay of statistical correlations already discussed in the stationary case.\par
It is interesting to study this self-similar decay more deeply in the framework of the PDF equation \eqref{eq:pdfveliso}. For decaying turbulence, it is particularly useful to employ the method of characteristics. Let $V(t,v_0)$ denote a characteristic curve which starts from $v_0$, i.e. $V(t_0,v_0)=v_0$. The method of characteristics then yields the following equations:
\begin{align}\label{eq:characteristics}
  \frac{\mathrm{d}}{\mathrm{d} t} V(t,v_0) &= \left[ \Pi(v,t)+\Lambda(v,t) \right]_{v=V(t,v_0)}, \\
  \frac{\mathrm{d}}{\mathrm{d} t} \tilde f(V(t,v_0);t) &= \left[ -\frac{\partial}{\partial v} \big( \Pi(v,t)+\Lambda(v,t) \big) \right]_{v=V(t,v_0)} \, \tilde f(V(t,v_0);t) ,
\end{align}
of which the latter is easily integrated. We obtain the evolution of $\tilde f$ along the characteristic curves:
\begin{equation}
  \tilde f(V(t,v_0);t) = \tilde f(v_0 ; t_0) \, \exp \left[ -\int_{t_0}^t\mathrm{d}t' \, \left[ \frac{\partial}{\partial v} \big( \Pi(v,t')+\Lambda(v,t') \big)\right]_{v=V(t',v_0)} \right] .
\end{equation}
Of course, we are more interested in the temporal evolution of $f(v;t)$ instead of $f(V(t,v_0);t)$. This mapping can be achieved with the inverse function of $V(t,v_0)$, which is defined by $V^{-1}(t,V(t,v_0)) = v_0$. Thus, the temporal evolution of the PDF of the magnitude of velocity is given by $\tilde f(v;t) = \tilde f(V(t,V^{-1}(t,v));t)$, such that we obtain
\begin{equation}
  \tilde f(v;t) =  \tilde f(V^{-1}(t,v) ; t_0) \, \exp \left[ -\int_{t_0}^t\mathrm{d}t' \, \left[ \frac{\partial}{\partial v'} \big( \Pi(v',t')+\Lambda(v',t') \big)\right]_{v'=V(t',V^{-1}(t,v))} \right] .
\end{equation}
The interesting fact about this result now is that we obtain self-similar solutions of the form \eqref{eq:selfsimilarpdf} if and only if the characteristic curves are of the form
\begin{equation}
  V(t,v_0) = \frac{\sigma(t)}{\sigma(t_0)} v_0,
\end{equation}
which due to \eqref{eq:characteristics} gives rise to the relation
\begin{equation}\label{eq:selfsim}
  \Pi(v,t)+\Lambda(v,t) = v \, \frac{\mathrm{d}}{\mathrm{d} t}\ln\left[ \frac{\sigma(t)}{\sigma(t_0)} \right] = -\frac{1}{2} \frac{\langle \varepsilon \rangle(t)}{E_{kin}(t)} \, v .
\end{equation}
This means that we only obtain self-similar solutions if the sum of the conditional averages related to the pressure gradient and the Laplacian of the velocity is a linear function of $v$, with a negative slope proportional to the ratio of the rate of energy dissipation and the kinetic energy. Note that this non-trivial relation is a direct consequence of the observation of a self-similar regime of the velocity PDF, no matter whether it is Gaussian or not. The relation is also checked with our numerical results with some examples shown in figure \ref{fig:selfsim}. Apart from the beginning of the decay phase, the analytical relation \eqref{eq:selfsim} is in very good agreement with measured conditional averages. \par
To characterize the statistics of decaying turbulence further, we study the temporal evolution of the kinetic energy and the rate of energy dissipation. The decay rate of energy is a central question in turbulence research, which still remains a point of debate. We refer the reader to \cite{karman38prs,kolmogorov41dan,batchelor48prs,saffman67pof,george92pfa} for a detailed account of the theoretical aspects of this question as well as some classical experimental results. While a power-law decay of the kinetic energy is widely accepted, the prediction for the precise numeric value of the decay exponent varies. Detailed numerical investigations on this have been presented recently in \cite[]{ishida06jfm,perot11aip}. Apart from the dependence of the decay exponent on some fundamental theoretical issues concerning the Loitsyansky integral, it has to be assumed that it depends on properties of the initial conditions \cite[]{george92pfa,ishida06jfm}. As we have not investigated a larger class of initial conditions, the following results shall characterize the simulations rather than making a point for or against some of the cited theories.\par
The kinetic energy for a single run is shown in figure \ref{fig:decayenergy}. After a short transient period of about $1T$ ($T$ is the large-eddy turnover time of the initial condition), the kinetic energy decays algebraically with an exponent close to $1.4$, i.e. $E_{kin}(t) \sim t^{-1.4}$, which is in good agreement with the results presented in \cite[]{ishida06jfm} and close to the Kolmogorov prediction of $E_{kin}(t) \sim t^{-10/7}$. Accordingly, the kinetic energy dissipation decays with an exponent close to $2.4$ in this regime, as shown in the inset of figure \ref{fig:decayenergy}. The self-similar decay together with an algebraic evolution of the variance indicates a particularly simple evolution in time. It is also interesting to note that the slight deviations from linearity of the conditional acceleration observed in figure \ref{fig:selfsim} lie within the short initial transient. It seems as if the turbulence needs a short relaxation time to switch from the forced regime to the decaying regime.\par

\subsection{Comparison with the Vorticity}\label{vorticity-section}

\begin{figure}
\centering\includegraphics[width=0.5\textwidth]{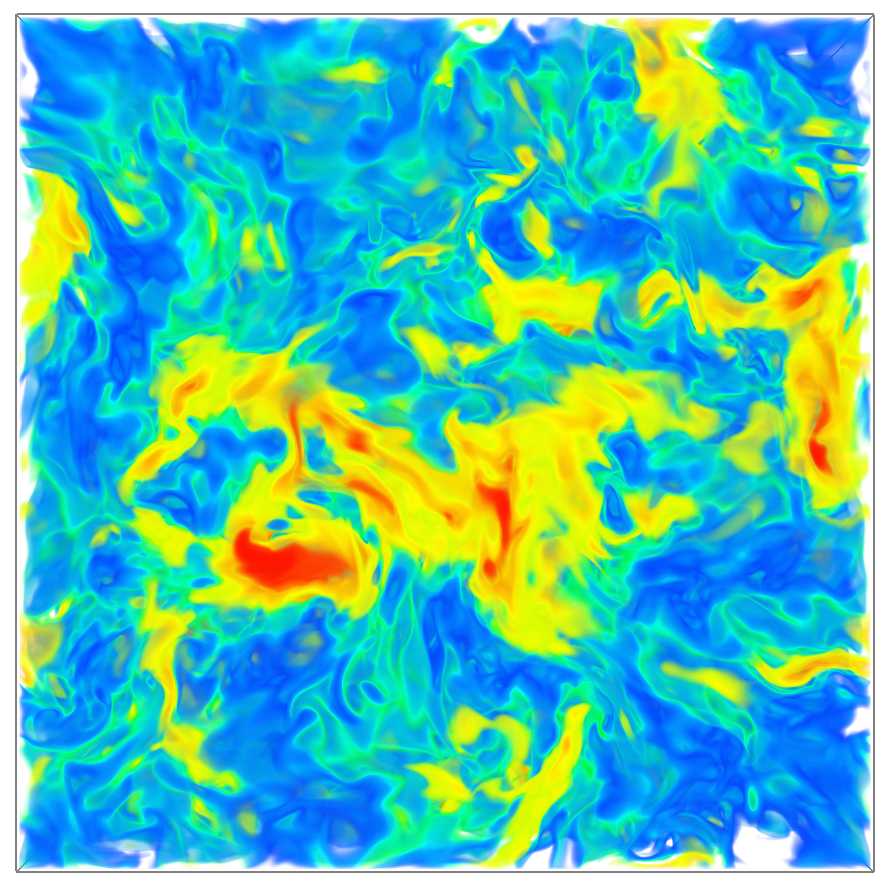}\includegraphics[width=0.5\textwidth]{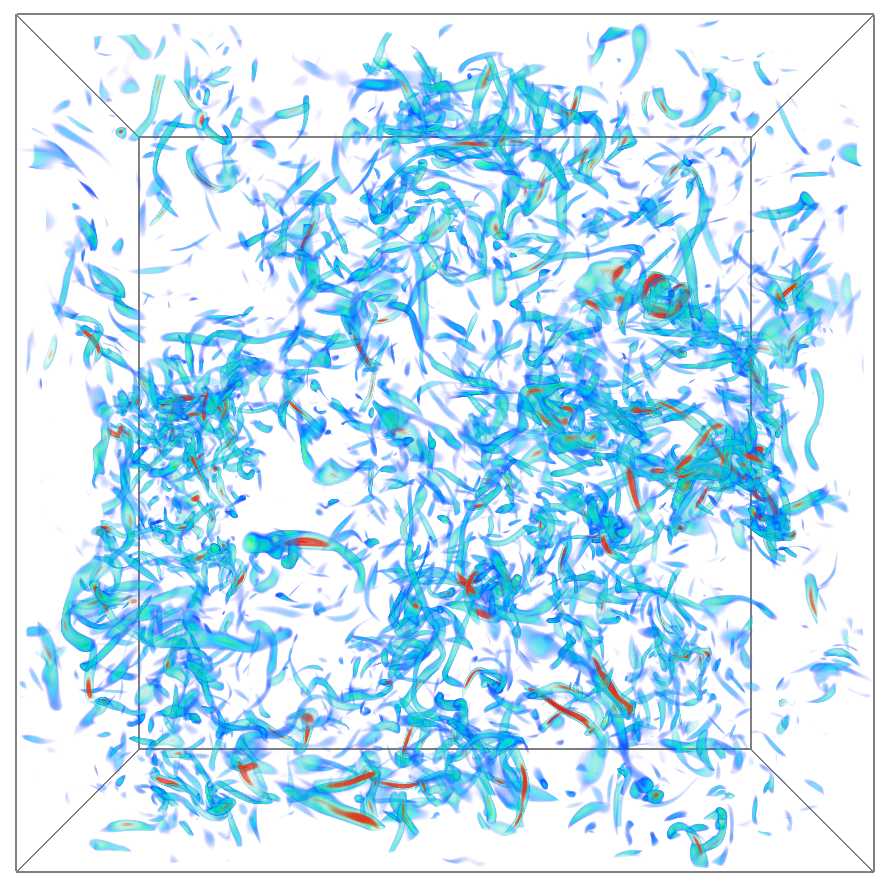}
\centering\includegraphics[width=0.5\textwidth]{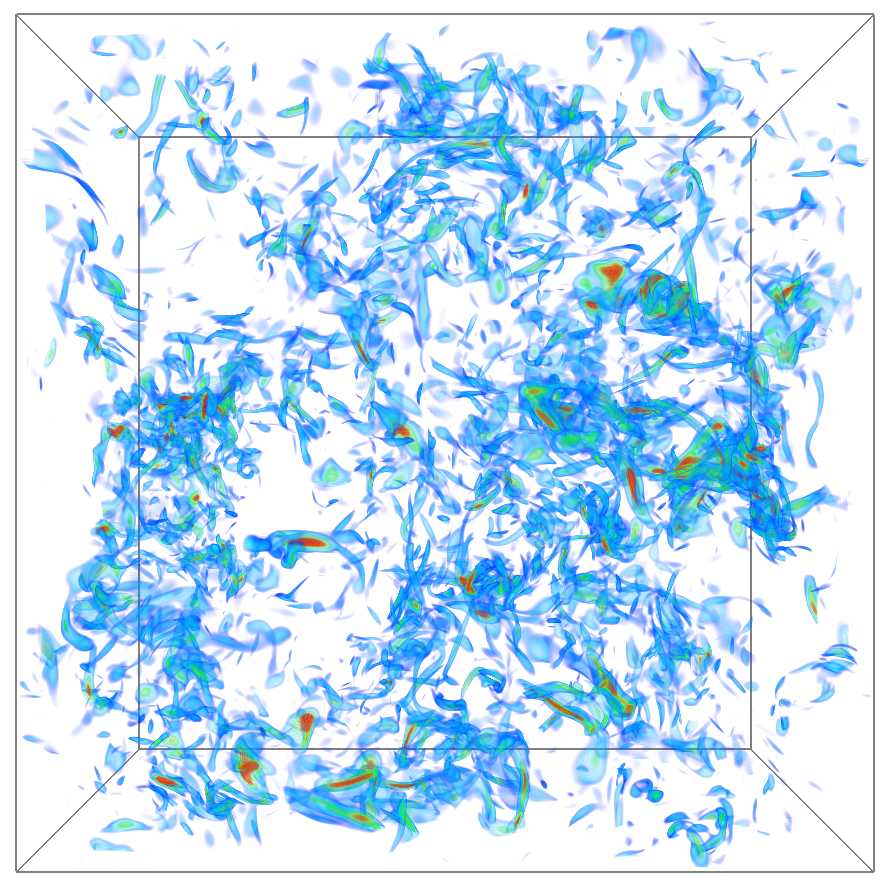}\includegraphics[width=0.5\textwidth]{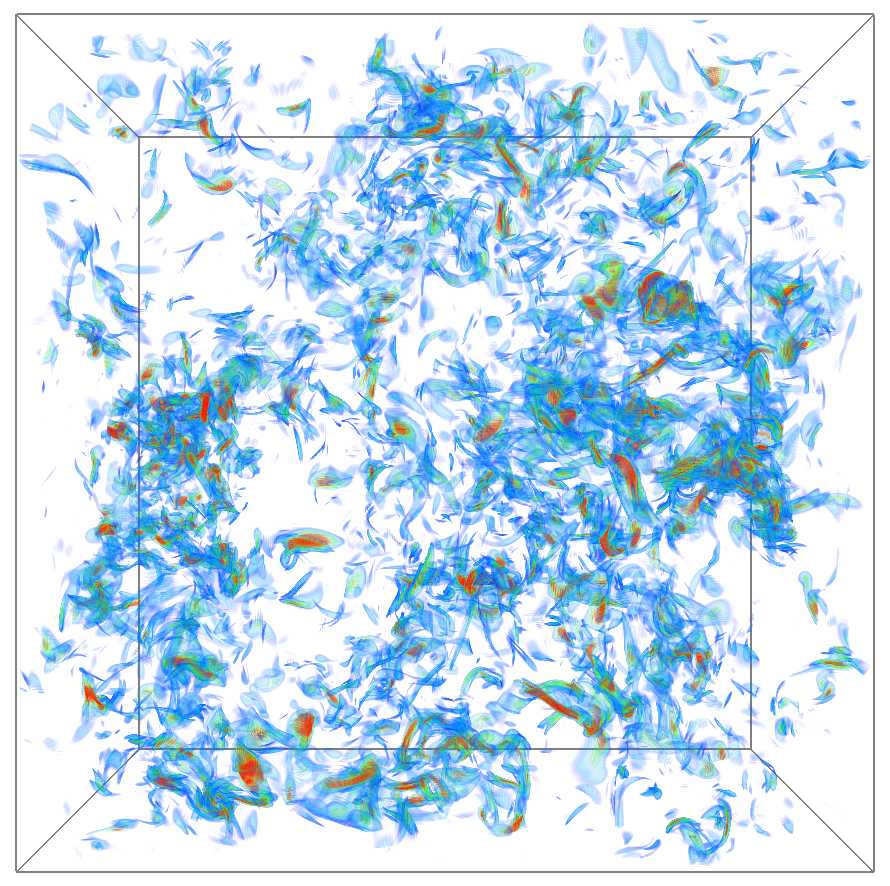}
\caption{Volume rendering of the magnitude of the velocity field (left top), the trace of the dissipation tensor $\nu\mathrm{Tr}(\mat{AA}^{\mathrm T})=\frac{1}{2}(\varepsilon+\nu \omega^2)$ (also known as pseudo-dissipation, left bottom), the magnitude of the vorticity field (right top) and the trace of the enstrophy dissipation tensor $\nu\mathrm{Tr}\left[ \bs{\nabla} \bs{\omega} (\bs{\nabla} \bs{\omega})^{\mathrm T} \right]$ (right bottom). The colours blue, green, yellow, red in that order indicate increasing amplitude. The vorticity and both dissipation tensors tend to organize into entangled small-scale structures forming a complicated global structure. The velocity appears more unstructured but displays long-range correlations. The visualizations have been produced with VAPOR (\texttt{www.vapor.ucar.edu}).}
\label{fig:vortvel}
\end{figure}

To demonstrate that some of the results are unique features of the velocity statistics, we draw a comparison with the statistics of the turbulent vorticity, which has been studied more extensively in \cite[]{wilczek09pre}. A first glance at snapshots of the fields reveals a fundamental difference between velocity fields and vorticity fields as shown in figure \ref{fig:vortvel}. While the vorticity tends to be organized in small filamentary structures, the velocity seems to be more structureless and smeared out. Recall that the velocity may be calculated from the vorticity with the help of the Biot--Savart law, which may be interpreted as some kind of a smoothing filter. On the level of the single-point statistics, the most striking difference is that the vorticity displays a highly non-Gaussian PDF with pronounced tails, which is closely related to the coherent structures present in the flow. Moreover, as for example reasoned in \cite[]{tennekes83book}, in the case of the turbulent vorticity the average enstrophy production and enstrophy dissipation tend to cancel, which means that the external forcing has a negligible effect on the statistics of the vorticity at high Reynolds numbers.

\begin{figure}
  \centering\includegraphics[width=0.5\textwidth]{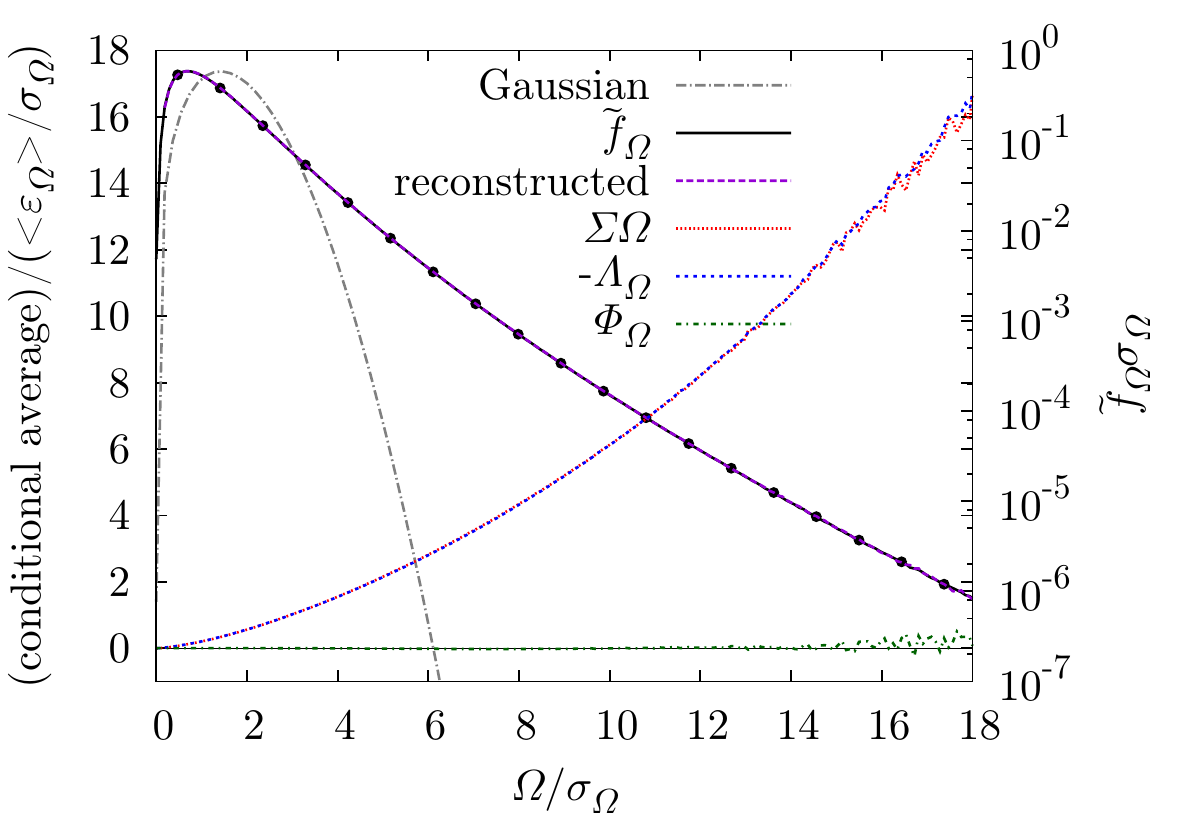}\includegraphics[width=0.5\textwidth]{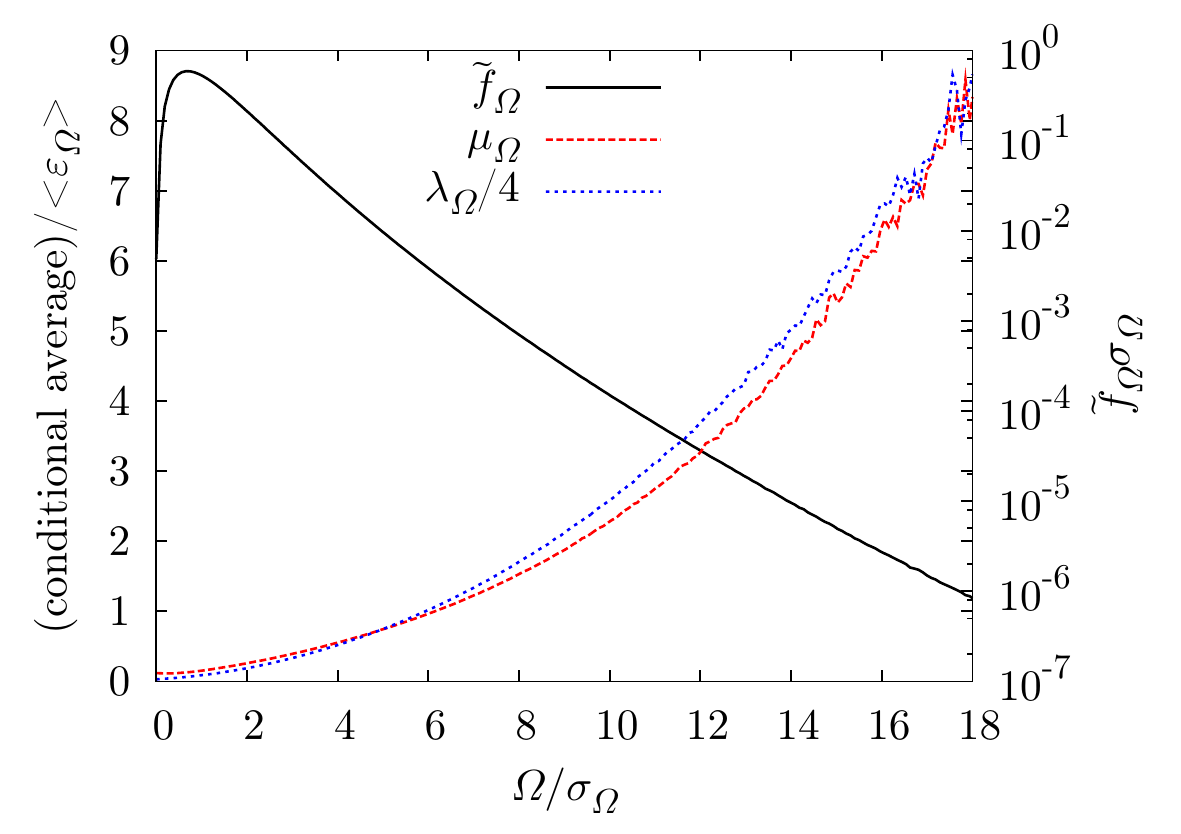}
  \caption{The PDF of the vorticity and the conditional averages which determine its shape. Note that the forcing term vanishes and therefore $\Lambda_\Omega$ and $\Sigma \, \Omega$ cancel. The abscissa is normalized by the standard deviation of the vorticity, whereas the ordinate is normalized using the average of $\varepsilon_\Omega=\nu/2\mathrm{Tr}\left( \left[ \bs{\nabla} \bs \omega+(\bs{\nabla} \bs \omega)^{\mathrm T}\right] ^2\right)$ and the standard deviation of the vorticity $\sigma_\Omega$. Note that not $\lambda_\Omega$ but $\lambda_\Omega/4$ is shown in the right figure.}
  \label{fig:everything_about_omega}
\end{figure}

The evolution equation for the vorticity PDF $f_{\Omega}(\bs \Omega; \bs x,t)$ can be derived analogously to the PDF equation for the velocity. For homogeneous turbulence, it takes the form
\begin{equation}
  \frac{\partial}{\partial t}f_{\Omega} = - \bs{\nabla}_{\bs \Omega}\cdot\left[ \left\langle  \mat{S} \bs \omega + \nu \Delta \bs \omega  + \bs{\nabla} \times \bs F\, \big | \bs \Omega \right\rangle  f_{\Omega} \right] \label{eq:evolution_liouville_omega}.
\end{equation}
Exploiting homogeneity, the diffusive term may also be rearranged, such that the PDF equation takes the form
\begin{equation}
  \frac{\partial}{\partial t}f_{\Omega} = -\bs{\nabla}_{\bs \Omega}\cdot \left[ \left\langle \mat{S} \bs \omega + \bs{\nabla} \times \bs F\, \big| \bs \Omega \right\rangle f_{\Omega} \right]
-\frac{\partial^2}{\partial \Omega_i \partial \Omega_j} \left\langle \nu \frac{\partial \omega_i}{\partial x_k} \frac{\partial \omega_j}{\partial x_k} \bigg| \bs \Omega \right\rangle \! f_{\Omega} .
\end{equation}
Hereby the conditional enstrophy dissipation tensor enters. Because of statistical isotropy the quantities in these equations take the form 
\begin{align}
 \left\langle \nu \Delta \bs \omega \big | \bs \Omega \right\rangle &= \Lambda_\Omega(\Omega) \hat{\bs \Omega}, 
\qquad \Lambda_\Omega(\Omega)=\left\langle \nu  \hat{\bs \omega} \cdot \Delta \bs \omega \big | \Omega \right\rangle,\\
 \left\langle \bs{\nabla} \times \bs F \big |  \bs \Omega \right\rangle &= \Phi_\Omega(\Omega) \hat{\bs \Omega}, 
\qquad \Phi_\Omega(\Omega)=\left\langle \hat{\bs \omega} \cdot (\bs{\nabla} \times \bs F) \big | \Omega \right\rangle,
\end{align}
\begin{equation}
 \left\langle S_{ij} \big| \bs \Omega \right\rangle = \frac{1}{2} \Sigma(\Omega) \left(3 \frac{\Omega_i \Omega_j}{\Omega^2} - \delta_{ij} \right),\qquad \Sigma(\Omega) = \left\langle \hat{\bs \omega} \mat{S} \hat{\bs \omega} \big| \Omega \right\rangle,
\end{equation}
\begin{equation}
 D_{\Omega;ij}(\bs \Omega)=\left\langle \nu \frac{\partial \omega_i}{\partial x_k} \frac{\partial \omega_j}{\partial x_k} \bigg | \bs \Omega \right  \rangle=\mu_\Omega(\Omega)\delta_{ij}+\left[ \lambda_\Omega(\Omega)-\mu_\Omega(\Omega)\right]\frac{\Omega_i \Omega_j}{\Omega^2},
\end{equation}
where $\hat{\bs \Omega}$ and $\hat{\bs \omega}$ are the unit vectors belonging to $\bs \Omega$ and $\bs \omega$.
The rate-of-strain tensor may, due to incompressibility, be characterized by a single scalar function $\Sigma(\Omega)$ related to the enstrophy production. The conditionally averaged vortex stretching term can then be expressed as
\begin{equation}
 \left\langle \mat{S} \bs\omega \big | \bs \Omega \right\rangle = \left\langle \mat{S} \big | \bs \Omega \right\rangle \bs\Omega=\Sigma(\Omega)\Omega\,\hat{\bs\Omega} .
\end{equation}
The eigenvalues of $\mat{D}_{\Omega}$ can be computed using relations analogous to \eqref{eq:mu_relation} and \eqref{eq:lambda_relation}. The isotropic forms of the evolution equations and the homogeneity relation (analogous to \eqref{eq:pdfveliso}--\eqref{eq:pdfvelhomoiso}) are
\begin{align}
 \frac{\partial}{\partial t} \tilde f_\Omega&=-\frac{\partial}{\partial \Omega} \left( \Sigma\,\Omega+\Lambda_\Omega+\Phi_\Omega \right) \tilde f_\Omega\label{eq:pdfveliso_omega}\\
0&=-\frac{\partial}{\partial \Omega} \left( \Lambda_\Omega+\frac{2\mu_\Omega}{\Omega} \right) \tilde f_\Omega + \frac{\partial^2}{\partial \Omega^2}  \lambda_\Omega  \tilde f_\Omega\label{eq:pdfhomoiso_omega}\\
 \frac{\partial}{\partial t} \tilde f_\Omega&=-\frac{\partial}{\partial \Omega} \left( \Sigma\,\Omega+\Phi_\Omega-\frac{2\mu_\Omega}{\Omega} \right) \tilde f_\Omega - \frac{\partial^2}{\partial \Omega^2} \lambda_\Omega \tilde f_\Omega\label{eq:pdfvelhomoiso_omega} .
\end{align}
Solving \eqref{eq:pdfhomoiso_omega} yields a relation between the PDF and the conditional averages
\begin{equation}\label{eq:homosol_omega}
  \tilde f_\Omega(\Omega;t)=\frac{{\cal N}}{\lambda_\Omega(\Omega,t)} \exp \int_{\Omega_0}^\Omega \mathrm{d}\Omega' \, \frac{ \Lambda_\Omega(\Omega',t)+\frac{2}{\Omega'}\mu_\Omega(\Omega',t)}{\lambda_\Omega(\Omega',t)}.
\end{equation}
The reconstruction of the PDF using this formula again works perfectly well, as shown in figure \ref{fig:everything_about_omega}.
For stationary turbulence, the probability current in \eqref{eq:pdfveliso_omega} has to vanish and therefore also the conditional average in \eqref{eq:evolution_liouville_omega}. However, it turns out that the effect of the external forcing may be neglected here, as can be seen from figure \ref{fig:everything_about_omega}. Therefore, the conditionally averaged vortex stretching term and the diffusive term tend to cancel almost identically. This means that the balance between enstrophy production and dissipation holds even under conditional averaging, which has already been shown in \cite[]{novikov93jfr,novikov94mpl,wilczek09pre}. As a consequence, we may regard the enstrophy production as an internal process, which is decoupled from the external forcing. This interpretation is consistent with the common assumption that the small-scale properties of turbulence become independent of the forcing mechanism given a sufficiently high Reynolds number. As demonstrated in the preceding paragraphs, the forcing term may not be neglected when determining the stationary PDF of the velocity. 

The conditional averages are shown in figure \ref{fig:everything_about_omega} together with the vorticity PDF. They display a strong dependence on $\Omega$, especially the eigenvalues which do not show a plateau for small values of $\Omega$ like in the case of velocity. An important observation here is that, in contrast to the velocity statistics $\lambda_\Omega\not\approx\mu_\Omega$, but rather $\lambda_\Omega\approx4\mu_\Omega$. This shows that the tensor $\mat{D}_\Omega$ depends on the direction of $\bs\Omega$, i.e. the enstrophy dissipation tensor $\nu\bs{\nabla}\bs{\omega}\left( \bs{\nabla}\bs{\omega} \right)^{\mathrm T}$ and the direction of the vorticity $\hat{\bs \Omega}$ are obviously correlated, in contrast to an almost vanishing correlation in the case of velocity. One may hypothesize that the simple scale-separation argument, which works in the case of the conditional dissipation tensor, breaks down for the conditional enstrophy dissipation tensor, because there is no clear scale separation between the vorticity and its gradients. This can be observed in figure \ref{fig:vortvel}, where a snapshot of the velocity, vorticity and the traces of the corresponding dissipation tensors are shown.
It appears that the fields of the vorticity and both dissipation tensors consist of structures of similar size, whereas the velocity field consists of fairly large patches. Furthermore, the structures\footnote{The vorticity field consists of tube-like structures, whereas the pseudo-dissipation field $\nu\mathrm{Tr}(\mat{AA}^{\mathrm T})=\frac{1}{2}(\varepsilon+\nu\omega^2)$ consists of structures, which are a superposition of sheet-like and tube-like structures.} of vorticity and the dissipation tensors are oriented in almost the same direction, but no obvious correlation of the \emph{orientation} of the structures of the velocity and the other fields is visible. However, the \emph{amplitudes} of the fields are correlated. Therefore it is understandable that $\mat{D}$ and $\mat{D}_\Omega$ both depend on the magnitude of the corresponding variable, but the dependence on the direction of the variable ($\hat{\bs v}$ and $\hat{\bs \Omega}$ respectively) is weak in the case of velocity and strong in the case of vorticity.

\begin{figure}
  \begin{center}
  \includegraphics[width=0.5\textwidth]{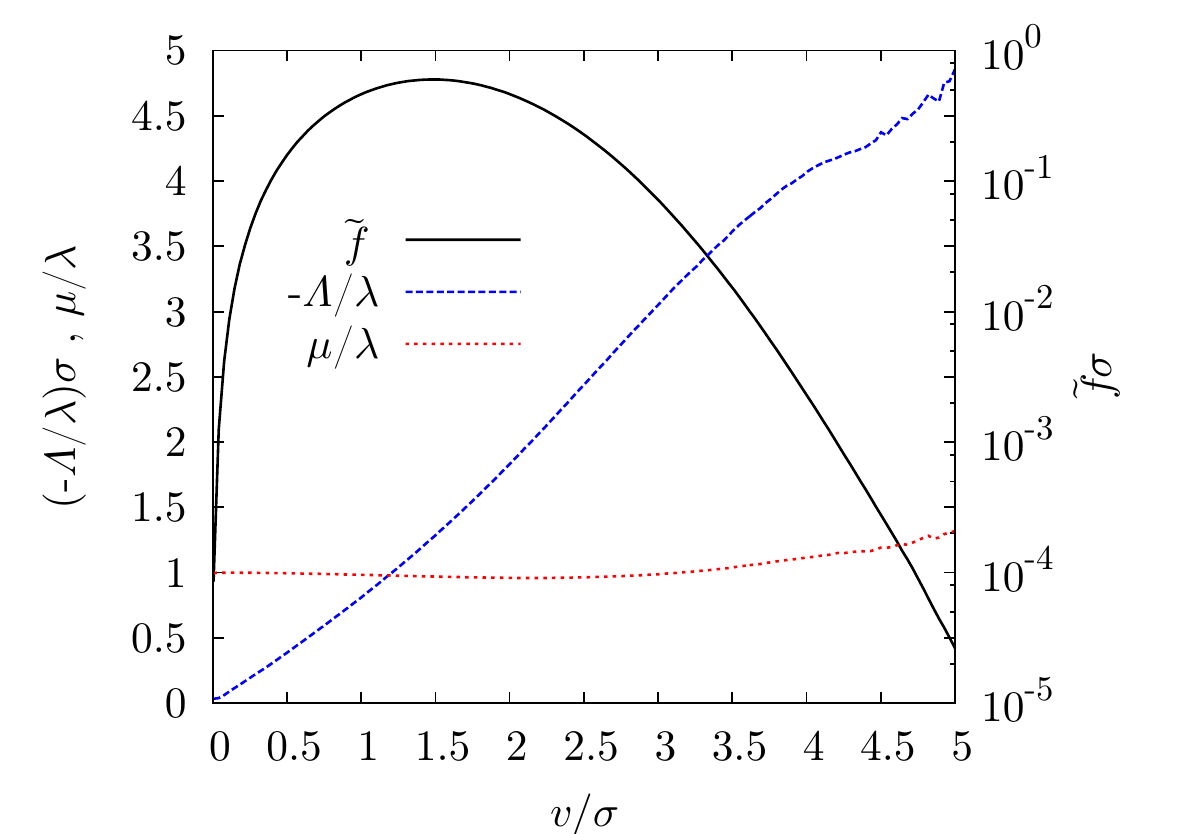}\includegraphics[width=0.5\textwidth]{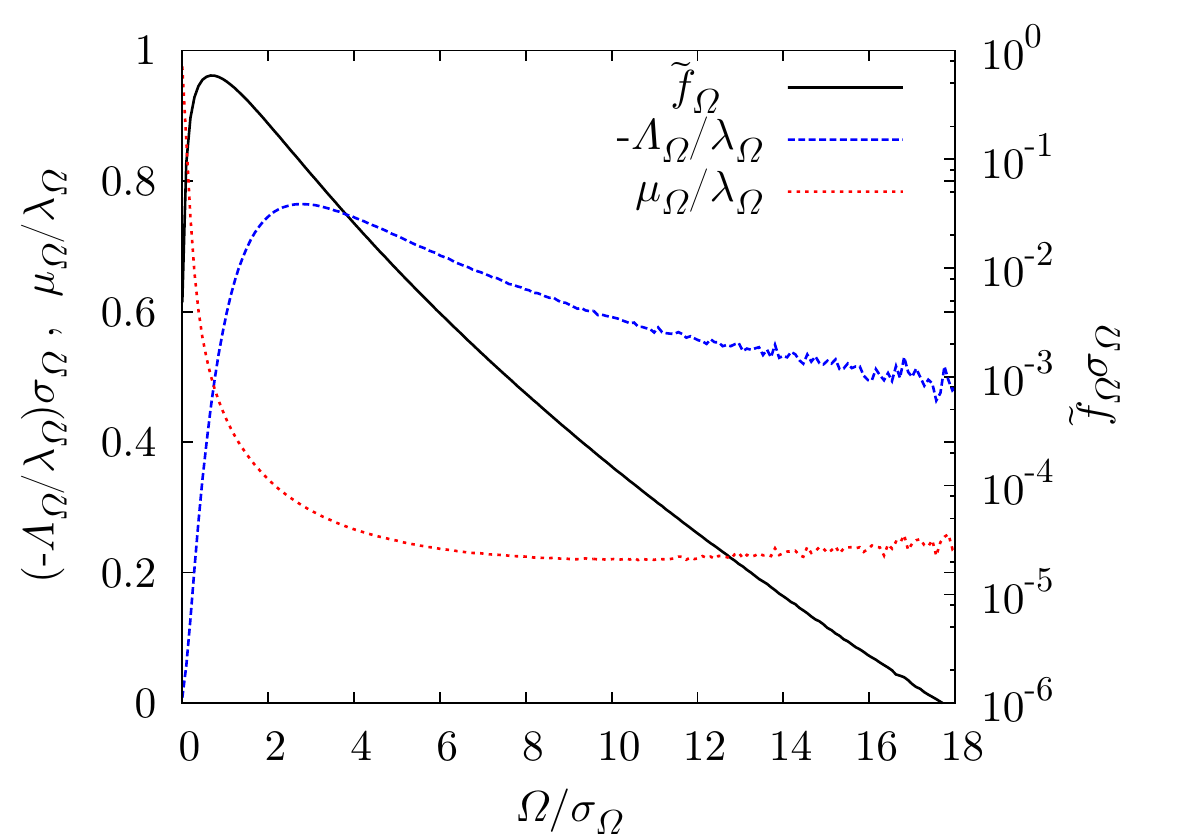}
  \end{center}
  \caption{The quotients $\Lambda/\lambda$ and $\mu/\lambda$ appearing in \eqref{eq:homosol} and \eqref{eq:homosol_omega} for the case of velocity (left) and vorticity (right).}
  \label{fig:integrands-analyzed}
\end{figure}

We have already noted that the arguments in \S\,\ref{sec:simple-closure}, which lead to a Gaussian PDF, could, in principle, also be applied to the vorticity. Figure \ref{fig:everything_about_omega} shows a comparison of $\tilde{f}_\Omega$ with an angle-integrated Gaussian. This exemplifies that such types of arguments can fail completely and therefore cannot be used as proofs of Gaussianity. To understand the reasons for a strongly non-Gaussian vorticity PDF we again examine the quotients $\Lambda_\Omega/\lambda_\Omega$ and $\mu_\Omega/\lambda_\Omega$, depicted in figure \ref{fig:integrands-analyzed}. We see that $\Lambda_\Omega/\lambda_\Omega$ is clearly not linear and has a complicated dependence on $\Omega$, showing that there is a complicated relation between vorticity diffusion and the enstrophy dissipation tensor. The quotient of the eigenvalues also shows a dependence on $\Omega$, corresponding to the fact that $\mu_\Omega\!\nsim\!\lambda_\Omega$. A detailed analysis shows that the complicated form of the vorticity PDF depends crucially on the non-trivial form of \emph{both} quotients, i.e. it cannot be attributed to one of them. Comparing the quotients for the case of velocity and vorticity, we see that there is much more non-trivial behaviour in the vorticity case. This leads to a highly non-Gaussian vorticity PDF, whereas the velocity PDF displays only moderate, but still significant, deviations from a Gaussian distribution. \par
In this section, we have examined the difference between the velocity and vorticity PDFs, using the conditional averages which determine these PDFs. We have seen that the approximate relations $\lambda\approx\mu$ and $\Lambda/\lambda\sim v$, which have been found for the velocity, are clearly violated in the case of vorticity. Overall, we can say that the conditional averages or the statistical dependences which determine the shape of the vorticity PDF are significantly more complicated for the vorticity and hence yield a highly non-Gaussian PDF.

\section{Summary and Conclusions}

To sum up, we studied the statistics of the single-point velocity PDF within the framework of the LMN hierarchy. Combined with conditional averaging and the use of statistical symmetries, this framework provides a clear-cut identification of the quantities which determine the details of the velocity PDF. The theory identifies the conditional diffusion of velocity and the conditional dissipation tensor in the homogeneous case as the central quantities of interest. In the case of stationary turbulence, the diffusive term is equivalently represented by the conditional pressure gradient and the external forcing. Exact expressions for the velocity PDF are presented in terms of these functions, demonstrating that the closure problem of turbulence may be expressed in terms of \textit{a priori} unknown statistical correlations of different dynamical contributions directly related to the Navier--Stokes equation.\par
In an analytical treatment, we suggest simple closure approximations that comply with the functional constraints on the conditional averages leading to Gaussian PDFs for both stationary and decaying turbulence. It turns out that the presented closure approximations contain the results recently derived in \cite[]{hosokawa08pre}, explaining which correlations have to be neglected to yield Gaussian statistics.\par
DNS simulations of stationary and decaying turbulence are then used to study deviations from Gaussianity. It turns out that the velocity PDF displays slight deviations from a Gaussian distribution with sub-Gaussian tails. A detailed investigation of the terms arising in the theoretical framework allows to investigate why only moderate deviations from Gaussianity occur in spite of pronounced correlations. Unlike suggested in recent theories, we show that the pressure contributions may not be neglected. Furthermore, the correlations of the velocity with the external forcing do not seem to be the main contributor to the deviations from Gaussianity. The conditional energy dissipation tensor turns out to display strong correlations for high values of the velocity, which contribute to the deviations. The numerical results additionally suggest that this tensor is approximately diagonal with identical eigenvalues related to the conditional rate of energy dissipation. The investigation of decaying turbulence gives similar results, indicating that forced and decaying turbulence do not differ in a fundamental way. Interestingly, a self-similar range with an algebraic decay of the kinetic energy has been observed, which considerably simplifies the description of the evolution of the PDF during the decay phase and implies a simple relation for the conditionally averaged right-hand side of the Navier--Stokes equation. \par 
To highlight the genuine features of the velocity statistics, some comparisons to the vorticity PDF are drawn. The general outcome here is that the vorticity is stronger correlated with the terms arising in the vorticity equation that determine the local dynamics such as enstrophy production and dissipation. The external forcing has a negligible effect as has been found in various other studies. The conditional enstrophy dissipation tensor cannot be assumed to be diagonal, in contrast to the conditional energy dissipation tensor, indicating the presence of non-negligible directional correlations. This fact may be related to the presence of coherent structures in the case of the vorticity and their absence in the case of the velocity; however, a concise description of this fact is still lacking.\par
In conclusion, the approach presented above gives a comprehensive characterization of the shape and evolution of the single-point velocity PDF. Although the introduction of conditional averages does not solve, but rather reformulates the closure problem of turbulence, valuable insights have been obtained highlighting the statistical correlations of the different dynamical influences and the velocity. Regarding a statistical theory of turbulence, it is interesting to see how PDFs may be expressed in terms of these influences. A theoretical derivation of the functional shape of these quantities remains a challenging task for the future.

\section*{Acknowledgments}
We thank O. Kamps, J. L\"ulff and F. Jenko for valuable discussions and careful reading of the manuscript as well as the Editor and one of the Referees for constructive input. Computational resources were granted within the project h0963 at the LRZ Munich.


\end{document}